\documentclass[fleqn,usenatbib]{mnras}

\usepackage{newtxtext,newtxmath}
\usepackage[T1]{fontenc}
\usepackage{ae,aecompl}

\usepackage{graphicx,color}% Including figure files
\usepackage{amsmath}% Advanced maths commands
\usepackage{amssymb}% Extra maths symbols
\usepackage{braket}

\usepackage{fancyvrb}
\usepackage{fancybox}
\newcommand{\ta}{{\mathrm{ta}}}
\title[Phase-space structure of CDM halos]{Phase-space structure of cold dark matter halos inside splashback: multi-stream flows and self-similar solution}

\author{Hiromu Sugiura}
\author[H. Sugiura et al.]{
\parbox{\textwidth}{
Hiromu Sugiura$^{1}$\thanks{E-mail: sugiura@tap.scphys.kyoto-u.ac.jp}, Takahiro Nishimichi$^{2,3}$, Yann Rasera$^{4}$, Atsushi Taruya$^{2,3}$
}
\vspace*{15pt} \\
$^{1}$Department of Physics, Kyoto University, Kyoto, 606-8502, Japan
\\
$^{2}$Center for Gravitational Physics, Yukawa Institute for Theoretical Physics, Kyoto University, Kyoto 606-8502, Japan
\\
$^{3}$Kavli Institute for the Physics and Mathematics of the Universe (WPI), The University of Tokyo Institutes for Advanced Study, \\
The University of Tokyo, 5-1-5 Kashiwanoha, Kashiwa, Chiba 277-8583, Japan
\\
$^{4}$LUTH, Observatoire de Paris, PSL Research University, CNRS, Universit\'e Paris Diderot, Sorbonne Paris Cit\'e 5 place Jules Janssen, 
\\
F-92195 Meudon, France
}
\begin{document}
\maketitle

\thisfancyput(14.8cm,0.5cm){\large{YITP-19-101}}

% Abstract of the paper
\begin{abstract}
Using the motion of accreting particles onto halos in cosmological $N$-body simulations, we study the radial phase-space structures of cold dark matter (CDM) halos. In CDM cosmology, formation of virialized halos generically produces radial caustics, followed by multi-stream flows of accreted dark matter inside the halos. In particular, the radius of the outermost caustic called the splashback radius exhibits a sharp drop in the slope of the density profile. Here, we focus on the multi-stream structure of CDM halos inside the splashback radius. To analyze this, we use and extend the SPARTA algorithm developed by Diemer. By tracking the particle trajectories accreting onto the halos, we count their number of apocenter passages, which is then used to reveal the multi-stream flows of the dark matter particles. The resultant multi-stream structure in radial phase space is compared with the prediction of the self-similar solution by Fillmore \& Goldreich for each halo. We find that $\sim30\%$ of the simulated halos satisfy our criteria to be regarded as being well fitted to the self-similar solution. The fitting parameters in the self-similar solution characterizes physical properties of the halos, including the mass accretion rate and the size of the outermost caustic (i.e., the splashback radius). We discuss in detail the correlation of these fitting parameters and other measures directly extracted from the $N$-body simulation.
\end{abstract}

\begin{keywords}
cosmology: theory -- dark matter -- methods: numerical
\end{keywords}

%%%%%%%%%%%%%%%%% BODY OF PAPER %%%%%%%%%%%%%%%%%%

%%%%%%%%%%%%%%%%%%%%%%%%%%%%%%%%%%%%%%%%%%%%%%%%%%%%%%%%%
%%%%%%%%%%%%%%%%%%%%%%%%%%%%%%%%%%%%%%%%%%%%%%%%%%%%%%%%%
\section{Introduction}
\label{sec:intro}
%%%%%%%%%%%%%%%%%%%%%%%%%%%%%%%%%%%%%%%%%%%%%%%%%%%%%%%%%
%%%%%%%%%%%%%%%%%%%%%%%%%%%%%%%%%%%%%%%%%%%%%%%%%%%%%%%%%

The concordant cosmological model, i.e., $\Lambda$CDM model, provides a simple picture of both the cosmic expansion and structure formation in the Universe with a minimal set of model parameters. The model consistently explains multiple cosmological observations, and the model parameters are measured precisely with the statistical error of a few percent level by the cosmic microwave background experiment, Planck \citep{Planck2018}.
According to this model, the large-scale matter inhomogeneities have evolved under the influence of gravity and cosmic expansion, starting with tiny density fluctuations which would have been generated in the early universe. An important ingredient of late-time structure formation driven by gravity is the cold dark matter (CDM),
which amount to more than $80\%$ of the matter components \citep{Peebles1982,Bond_etal1982,Blumenthal_etal1982}.
As it is named, the CDM was initially cold with negligibly small velocity dispersion, 
and behaved like dust fluid at the very early stage of structure formation.
Later, due to the attractive force of gravity, the CDM gradually accretes into overdense regions, and matter concentration grows. When the amplitude of the density contrast exceeds unity, the growth of fluctuations becomes nonlinear, finally ending up with the formation of self-gravitating bounded objects called dark matter halos through the collapse and virialization \citep{bt08}.
Since a sufficient amount of baryon has been accumulated by the gravitational potential well of dark matter after the recombination epoch,
the dark matter halo is an ideal site of galaxy and star formation,
and thus observationally important to probe the structure formation and cosmology.

Within the CDM paradigm, there have been numerous works to characterize the kinematical, dynamical, and statistical properties of dark matter halos. One important feature found in numerical simulations but not yet clearly understood is the cuspy density profile called the NFW profile \citep{nfw96}. Unlike naive theoretical expectations, the radially averaged density profile $\rho(r)$ near the halo center exhibits a shallow cusp, whose logarithmic slope, defined by $d\log\rho/d\ln r$, is larger than $-2$, mostly independent of cosmology and the size of halos. Another striking feature, also found in the cosmological $N$-body simulations, is the power-law nature of the pseudo phase-space density profile defined by $\rho(r)/\sigma^3(r)$, with $\sigma(r)$ being the velocity dispersion \citep{tn01, n10, l10}. The slope found in the simulations closely match the prediction of the Bertschinger's secondary infall model \citep{b85}, suggesting that the structure of halos is built up with continuous accretion flow and mergers. Yet, recalling the fact that the halos are not fully spherical but generically asymmetric with sizable amount of substructures, how such a simple picture can reconcile with the actual halo formation processes still remains unclear. Viewing the halo formation from the viewpoint of collisionless self-gravitating system, CDM halos generally have some memories of the initial condition, and due to its cold nature, unique and characteristic features appears manifest, in particular, in phase space. In fact, generic properties of the CDM halo mentioned above are linked with each other, and one expects that these are originated from the phase-space dynamics of the CDM halos. In this respect, the structural and statistical properties of the halos in phase space is worth for investigation, and there are thus numerous works along the line of this \citep[e.g.,][for recent works]{Drakos_etal2017,Halle_etal2019}. Also, a quantitative phase-space study would serve as a clue to discriminate CDM from non-standard dark matter scenarios, and in combination with observations, it may help to clarify the nature of dark matter \citep[e.g.,][]{Sikivie1997}.

To be more precise, the CDM inside halos is expected to have underwent shell crossing during the accretion, and the velocity at a given position gets multi-valued. On the other hand, the regions outside halos exhibits a single-stream flow, for which the velocity of accreting matter is uniquely determined as a function of position. Importantly, the collisionless and Hamiltonian nature of the system ensures that the phase-space density is conserved, and its topological structure remains unchanged. Thus, the single-stream flow should smoothly be connected to the multi-stream flow. Recently, \cite{dk14} pointed out that there is a significant deviation of the density profile from the NFW profile at the outskirt of halos, and this can happen exactly at the boundary between single- and multi-stream flow regions \citep{adc14}. In $N$-body simulation, the location of this boundary corresponds to the first apocenter of the accreting dark matter particles, particularly referred to as the splashback radius. Because of its clear manifestation, the observational prospects and the theoretical understanding of the splashback feature as a unique signature of the CDM paradigm have attracted much attention \citep{mdk15, more16, moremiyatake, shi16, busch17, sparta1, sparta2, adhikari18, okumura18, c18}.

It is theoretically expected that the splashback feature in the radial density profile appears more prominent in spherically symmetric halos, for which several self-similar solutions are known in the Einstein-de Sitter universe \citep[e.g.,][]{fg84,b85,wz92,ryden93,zb10,vmw11,lw11,alard13}. Because of the exact spherical symmetry, the density profile of self-similar solution exhibits apparent divergences called caustics at the apocenters of each flow of accreting matter. The outermost caustic particularly shows the most notable feature, and its location exactly corresponds to the splashback radius \citep{adc14}. \citet{shi16} used the self-similar solution by \citet{fg84} to give an analytical prediction of the splashback radius, and \citet{sparta2} found good agreement with numerical simulations. This suggests that the self-similar solution may capture the overall trends in the dynamics of accreting material on to CDM halos in simulations, and possibly those in the real universe if the CDM scenario is true, although it is very hard to imagine that spherically symmetric and isolated halo is realized in reality. In fact, even when starting from a nearly spherically symmetric initial condition, non-sphericity is rapidly developed due to the so-called radial-orbit instability \citep[e.g.,][]{bt08}, and a deviation from the top-hat spherical collapse model is significant \citep{suto16}. The resultant halo exhibits an elongated triaxial shape \citep[e.g.,][]{jingsuto02,suto16b}, rather different from the prediction of the self-similar solution \citep[e.g.,][]{macmillan06}. Nevertheless, the growth of halos and the evolved density profile are found to match the prediction of the self-similar solution. There are also several works advocating that taking spherical average, the phase-space structures of halos in $N$-body simulations resembles the spherical self-similar solutions \citep[e.g.,][]{bg91,henriksen97,mcgst06,v09,vw11,ddk13}. In these respects, it is still interesting to further clarify the similarities and differences between the self-similar solution and the full dynamics in $N$-body simulations in more quantitative manner. In particular, little work has focused on the multi-stream structure of CDM halos, and a detailed analysis from the phase space point-of-view has not yet been made. 

In this paper, we compare the phase-space structure of halos in a cosmological $N$-body simulation with those predicted from the self-similar solution, and try to clarify to what extent the multi-stream features agree between the two descriptions. Although the internal structures of halos are driven by the collisionless gravitational dynamics and thus the memory of initial condition should still remain preserved to some extent, generic properties of halos, including the universality in the density profile or the pseudo phase-space density scaling, are built up along the halo formation processes. In this respect, a phase-space comparison with self-similar solution would give a useful guideline or hint to understand how the generic features emerge and what environment-dependent features remain especially in the internal halo structures. In doing so, the statistical analysis using a large number of halos is important, and in this paper, we will make a detailed comparison of radial phase-space structures with self-similar solution for massive halos found in an $N$-body simulation. A crucial point in our present work is to extract different streams in each halo to reveal the multi-streaming structure. For this purpose, we adopt and extend the SPARTA algorithm by \citet{sparta1}. In short, using a number of output data at different redshifts, we keep track of the trajectories of dark matter particles around the halos, and count the number of apocenter passages for each dark matter particle. Sorting out all the particles around a halo with the number of apocenter passages, we can visualize, in phase space, each stream line of multi-stream flows inside the halo.

This paper is organized as follows.
In Sec.~\ref{sec:self-similar}, we present a brief review of the spherical self-similar solutions. Sec.~\ref{sec:method} describes the method that we adopt to analyze the dark matter halos identified in an $N$-body simulation. We show our results in Sec~ \ref{sec:result} and discuss its implication in Sec.~\ref{sec:discussion}.
Our conclusion from the analysis and further discussions are finally presented in Sec.~\ref{sec:conclusion}.

%%%%%%%%%%%%%%%%%%%%%%%%%%%%%%%%%%%%%%%%%%%%%%%%%%%%%%%%%
\begin{figure*}
  \begin{minipage}[b]{0.3\linewidth}
    \centering
    \includegraphics[clip,height=50truemm]{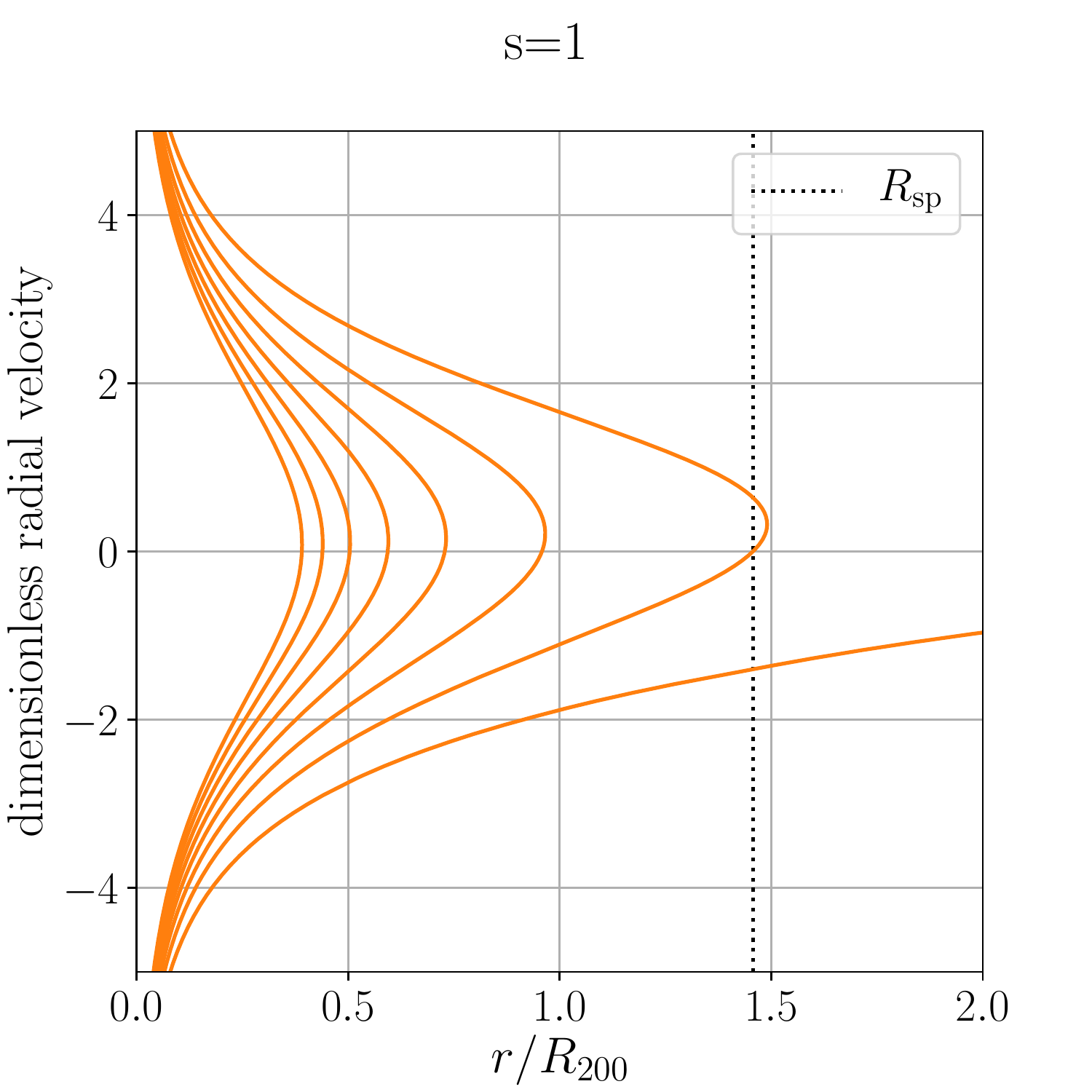}
  \end{minipage}
  \begin{minipage}[b]{0.3\linewidth}
    \centering
    \includegraphics[clip,height=50truemm]{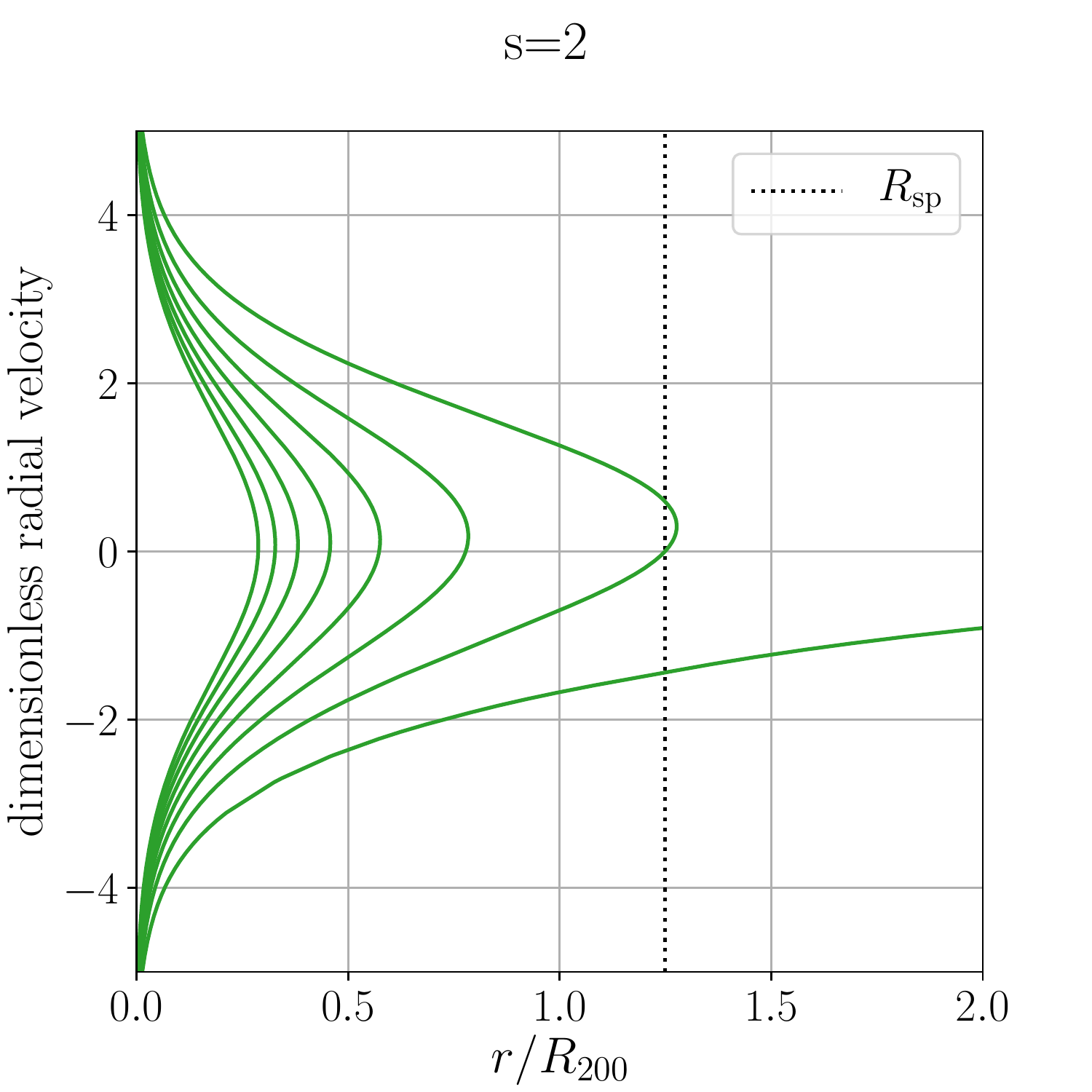}
  \end{minipage}
  \begin{minipage}[b]{0.3\linewidth}
    \centering
    \includegraphics[clip,height=50truemm]{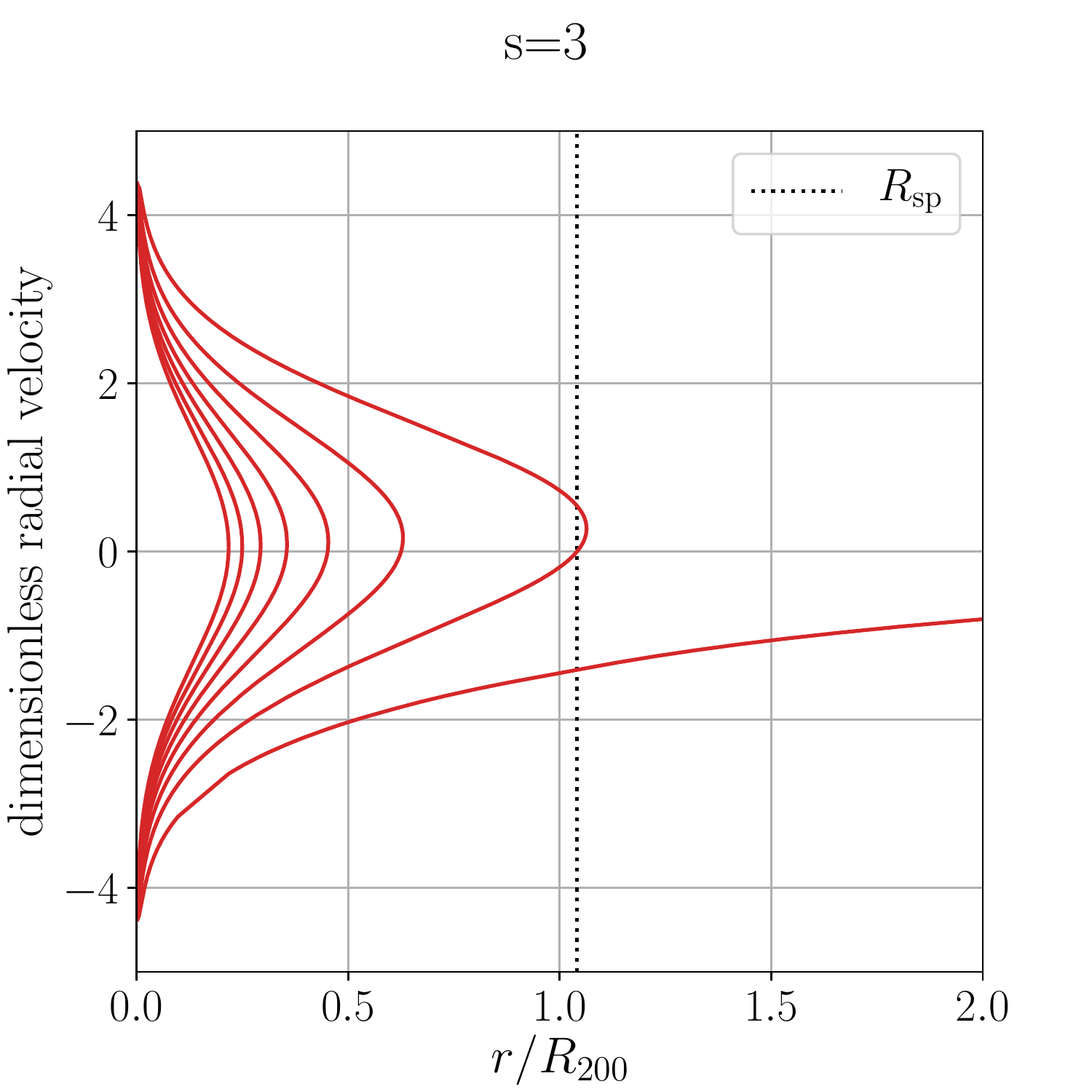}
  \end{minipage}
  \caption{Phase-space portraits of the self-similar solution for $s = 1$ (left), $2$ (middle) and $3$ (right). The horizontal axis represents the radial position normalized by the radius $R_{200}$, at which the mean overdensity of the halo reaches $200$ times the background mass density. The vertical axis means the dimensionless velocity, $\tau^{1-\beta}(d\lambda/d\tau)$. 
  The plot shows the trajectories up to the seventh apocenter passages. The vertical dotted lines indicates the splashback radius $R_\mathrm{sp}$ at which the trajectory crosses the zero-velocity line. }
  \label{fig-selfsimilar-solution}
\end{figure*}
%%%%%%%%%%%%%%%%%%%%%%%%%%%%%%%%%%%%%%%%%%%%%%%%%%%%%%%%%

%%%%%%%%%%%%%%%%%%%%%%%%%%%%%%%%%%%%%%%%%%%%%%%%%%%%%%%%%
%%%%%%%%%%%%%%%%%%%%%%%%%%%%%%%%%%%%%%%%%%%%%%%%%%%%%%%%%
\section{Spherical self-similar solutions} 
\label{sec:self-similar}
%%%%%%%%%%%%%%%%%%%%%%%%%%%%%%%%%%%%%%%%%%%%%%%%%%%%%%%%%
%%%%%%%%%%%%%%%%%%%%%%%%%%%%%%%%%%%%%%%%%%%%%%%%%%%%%%%%%

In this section, we present a brief review of the self-similar solution described by \citet{fg84} and \citet{b85}.

Consider a spherically symmetric density contrast in the Einstein-de Sitter universe, with surrounding materials stationary accreting toward the center. The dynamics of such a system is described by a collection of spherical shells moving radially. Although all the shells move outward in the radial direction according to the Hubble-Lema\^itre law at first, as increasing the central density due to the gravitational growth, the motion of surrounding shells ceases to follow the cosmic expansion, and start to infall into the central region \citep[e.g.,][]{gg72,gunn77,peebles80}. The time of this critical point is referred to as the turn-around time $t_{\rm ta}$, and the physical size/radius of the shell at that time is called the turn-around radius $r_{\rm ta}$, which is given as the function of $t_{\rm ta}$. Since each shell has different turn-around time, the properties of the system can be characterized by a family of the shell radii parametrized by $t_{\rm ta}$, hence we denote it by $r ( t, t_\ta )$.

Imposing the self-similarity, the function $r ( t, t_\ta )$ can be written in the form as
%%%%%%%%%%%%%%%%%%%%%%%%%%%%%%%%%%%%%%%%%%%%%%%%%%%%%%%%%%%%%%%%%%%%%%%%%%%%%%%%%%
\begin{align}
  r ( t, t_\ta ) = r_\ta ( t_\ta ) \lambda ( t / t_\ta ),
  \label{eq:radius_shell}
\end{align}
%%%%%%%%%%%%%%%%%%%%%%%%%%%%%%%%%%%%%%%%%%%%%%%%%%%%%%%%%%%%%%%%%%%%%%%%%%%%%%%%%%
where $\lambda$ is a dimensionless quantity. The functional form of $r_\ta ( t_\ta )$ depends on the initial condition.
Assuming a power-law for the initial density contrast given by $\delta_\mathrm{i} \propto r^{- 3/s}$, we have
%%%%%%%%%%%%%%%%%%%%%%%%%%%%%%%%%%%%%%%%%%%%%%%%%%%%%%%%%%%%%%%%%%%%%%%%%%%%%%%%%%
\begin{align}
  r_\ta ( t_\ta ) \propto t_\ta^\beta , \ \ \beta = \frac{ 2 }{ 3 } + \frac{ 2 }{ 9 } s .
\end{align}
%%%%%%%%%%%%%%%%%%%%%%%%%%%%%%%%%%%%%%%%%%%%%%%%%%%%%%%%%%%%%%%%%%%%%%%%%%%%%%%%%%
The parameter, $s$, introduced above is related to the mass accretion rate, and it is expressed as $s=d \ln M_\ta / d \ln a$, where $M_\ta$ is the enclosed mass within $r_\ta$ at the turn around epoch $t_\ta$ and $a \propto t^{2/3}$ is the scale factor of the Universe \citep{adc14,shi16}\footnote{In \citet{fg84}, they use $\epsilon=1/s$, instead of $s$.}. Note that this parameter $s$ fully determines the asymptotic inner-slope of the density profile, $\gamma \equiv d \ln \rho / d \ln r$, through
%%%%%%%%%%%%%%%%%%%%%%%%%%%%%%%%%%%%%%%%%%%%%%%%%%%%%%%%%%%%%%%%%%%%%%%%%%%%%%%%%%
\begin{align}
  \gamma = - \frac{ 9 }{ 3 + s } \ \ \text{for} \ s \leq \frac{3}{2} , \ \
  \gamma = - 2 \ \ \text{for} \ s \geq \frac{3}{2} .
  \label{eq:asymptotic_slope}
\end{align}
%%%%%%%%%%%%%%%%%%%%%%%%%%%%%%%%%%%%%%%%%%%%%%%%%%%%%%%%%%%%%%%%%%%%%%%%%%%%%%%%%%
With these setup, the solution in the special case with $s=1$ corresponds to the self-similar solution of the collisionless secondary infall by \citet{b85}.

In Eq.~(\ref{eq:radius_shell}), the function $\lambda ( \tau )$ is obtained by solving the equation of motion for shells:
%%%%%%%%%%%%%%%%%%%%%%%%%%%%%%%%%%%%%%%%%%%%%%%%%%%%%%%%%%%%%%%%%%%%%%%%%%%%%%%%%%
\begin{align}
  \frac{ d^2 r }{ d t^2 } = - \frac{ G M }{ r^2 } ,
  \label{eq:enclosed_mass}
\end{align}
%%%%%%%%%%%%%%%%%%%%%%%%%%%%%%%%%%%%%%%%%%%%%%%%%%%%%%%%%%%%%%%%%%%%%%%%%%%%%%%%%%
where $M$ is the mass enclosed by the shell. Under the assumption of self-similarity,
this equation is reduced to the non-dimensional form \citep{fg84}:
%%%%%%%%%%%%%%%%%%%%%%%%%%%%%%%%%%%%%%%%%%%%%%%%%%%%%%%%%%%%%%%%%%%%%%%%%%%%%%%%%
\begin{align}
  \frac{ d^2 \lambda }{ d \tau^2 } = - \frac{ \pi^2 }{ 8 } \frac{ \tau^{2s/3} }{ \lambda^2 } \mathcal{M} ( \lambda / \tau^\beta ) .
  \label{eom ss}
\end{align}
%%%%%%%%%%%%%%%%%%%%%%%%%%%%%%%%%%%%%%%%%%%%%%%%%%%%%%%%%%%%%%%%%%%%%%%%%%%%%%%%%
Here, the function $\mathcal{M} ( \xi )$ is a non-dimensional mass variable corresponding to the enclosed mass $M$ in Eq.~(\ref{eq:enclosed_mass}), i.e., the mass profile normalized by the turnaround mass, given in the integral form:
%%%%%%%%%%%%%%%%%%%%%%%%%%%%%%%%%%%%%%%%%%%%%%%%%%%%%%%%%%%%%%%%%%%%%%%%%%%%%%%%%
\begin{align}
  \mathcal{M} ( \xi ) = \frac{ 2 s }{ 3 }
  \int_1^\infty \Theta \left[ \xi - \frac{ \lambda ( \tau' ) }{ \tau'^\beta } \right] \frac{ d \tau' }{ \tau'^{1+2s/3} },
  \label{eq:mass_profile}
\end{align}
%%%%%%%%%%%%%%%%%%%%%%%%%%%%%%%%%%%%%%%%%%%%%%%%%%%%%%%%%%%%%%%%%%%%%%%%%%%%%%%%%
where $\Theta ( x )$ is the Heaviside step function. Thus, Eq.~(\ref{eom ss}) is the integro-differential equation, which has to be solved numerically based on an iterative method. That is, first we take an initial-guess for the mass profile and solve the equation of motion. We set $\mathcal{M} ( \xi ) = \xi$ as our simple initial guess. The solution for $\lambda$ obtained at the first trial is then used to estimate $\mathcal{M}$ through Eq.~(\ref{eq:mass_profile}), which will be next used to solve Eq.~(\ref{eom ss}) in the second trial. We repeat this procedure until the radial positions of the first five apocenters (i.e., the position at which $d\lambda/d\tau=0$ is satisfied) are converged well within the accuracy of 0.1\%. In solving Eq.~(\ref{eom ss}) in practice, we need to introduce a small angular momentum to avoid the singular behavior at $\lambda=0$ \citep{b85, mohayaee06}. This alters the solution near the center, and we adjust the angular momentum so that its impacts on the locations of first five apocenters are less than $0.1\,\%$. We calculated the self-similar solutions in the parameter range $0.1 \leq s \leq 9$.

The numerical solution of $\lambda$, given as a function of $\tau=t/t_{\rm ta}$, describes the motion of a single shell specified by a value of $t_{\rm ta}$. If we instead fix $t$ and draw $\lambda$ as a function of $t_\ta$,  it can give a snapshot of the distribution of a family of shells parameterized by $t_{\rm ta}$.  In other words, at a given time $t$, the function $r ( t, t_\ta )$ and its time derivative $\partial r/\partial t$, plotted as function of $t_{\rm ta}$, forms radial phase-space trajectories for a collection of shells:
%%%%%%%%%%%%%%%%%%%%%%%%%%%%%%%%%%%%%%%%%%%%%%%%%%%%%%%%%%%%%%%%%%%%%%%%%%%%%%%%%
\begin{align}
  \begin{aligned}
    & \left( r ( t, t_\ta ) , \ \frac{ \partial r }{ \partial t } ( t, t_\ta ) \right)_{t_\ta \in ( 0, t ]} \\
    = &\left( r_\ta ( t ) \tau^{-\beta} \lambda ( \tau ) , \
    \frac{ r_\ta ( t ) }{ t } \tau^{1-\beta} \frac{ d \lambda }{ d \tau }( \tau ) \right)_{\tau \in [ 1, \infty )}
  \end{aligned}
  \label{phase-distribution}
\end{align}
%%%%%%%%%%%%%%%%%%%%%%%%%%%%%%%%%%%%%%%%%%%%%%%%%%%%%%%%%%%%%%%%%%%%%%%%%%%%%%%%%

Fig.~\ref{fig-selfsimilar-solution} shows the snapshots of the self-similar solution for specific values of the parameter, $s = 1$ (left), $2$ (middle) and $3$ (right). Here, the horizontal axis in each panel is normalized by the radius $R_{200}$, within which the mean overdensity exceeds $200$ times the background mass density (or equivalently the critical density in the Einstein-de Sitter universe), and the vertical axis represents the dimensionless velocity, i.e., $\tau^{1-\beta}(d\lambda/d\tau)$.
The size of halos is supposed to be characterized by the radius $R_{200}$ roughly corresponding to the virial radius, but the actual size/region where the multi-stream flow can be extended out to a larger radius depending on the mass accretion rate parameter $s$. Overall, the size of the multi-stream region tends to get compressed as increasing $s$. In this respect, the so-called splashback radius, $r_\mathrm{sp}$, as indicated by the vertical dotted line in Fig.~\ref{fig-selfsimilar-solution}, provides a more appropriate definition of the size of a halo. Here, the location of the splashback radius is determined by the outermost location that satisfies the condition $(\partial r /\partial t )_{r_{\rm sp}}=0$. Note that in general, the location of the outermost caustic, defined by $dv/dr=0$, does not precisely coincide with the splashback radius defined here, although several works have used the outermost caustic as the boundary of a halo, which can be clearly seen from the sudden change in the slope of the radial density profile \citep[e.g., ][]{dk14,mdk15}. The reason why the locations of the caustic and the apocenter are different in phase space basically follows from the stationary mass accretion. That is, looking at the motion of shells, we see that the apocenter radius for each shell becomes gradually increasing in time $t_\ta$ due to the continuous mass growth at the center. Then, viewing a collection of shells in phase space at a given time $t$, we can find a small segment in the flow line that have a positive radial velocity, i.e., the shells which have not yet experienced an apocenter passage, but have a radial coordinate larger than the preceding shell which has just undergone an apocenter passage $(v=0)$, i.e., the splashback radius. In general, the radial location of the caustic tends to be larger than that of the apocenter in the presence of mass accretion.

%%%%%%%%%%%%%%%%%%%%%%%%%%%%%%%%%%%%%%%%%%%%%%%%%%%%%%%%%%%%%%%%%%%%%%%%%%%%%%%%%
%%%%%%%%%%%%%%%%%%%%%%%%%%%%%%%%%%%%%%%%%%%%%%%%%%%%%%%%%%%%%%%%%%%%%%%%%%%%%%%%%
\begin{figure*}
  \centering
  \includegraphics[clip,width=140truemm]{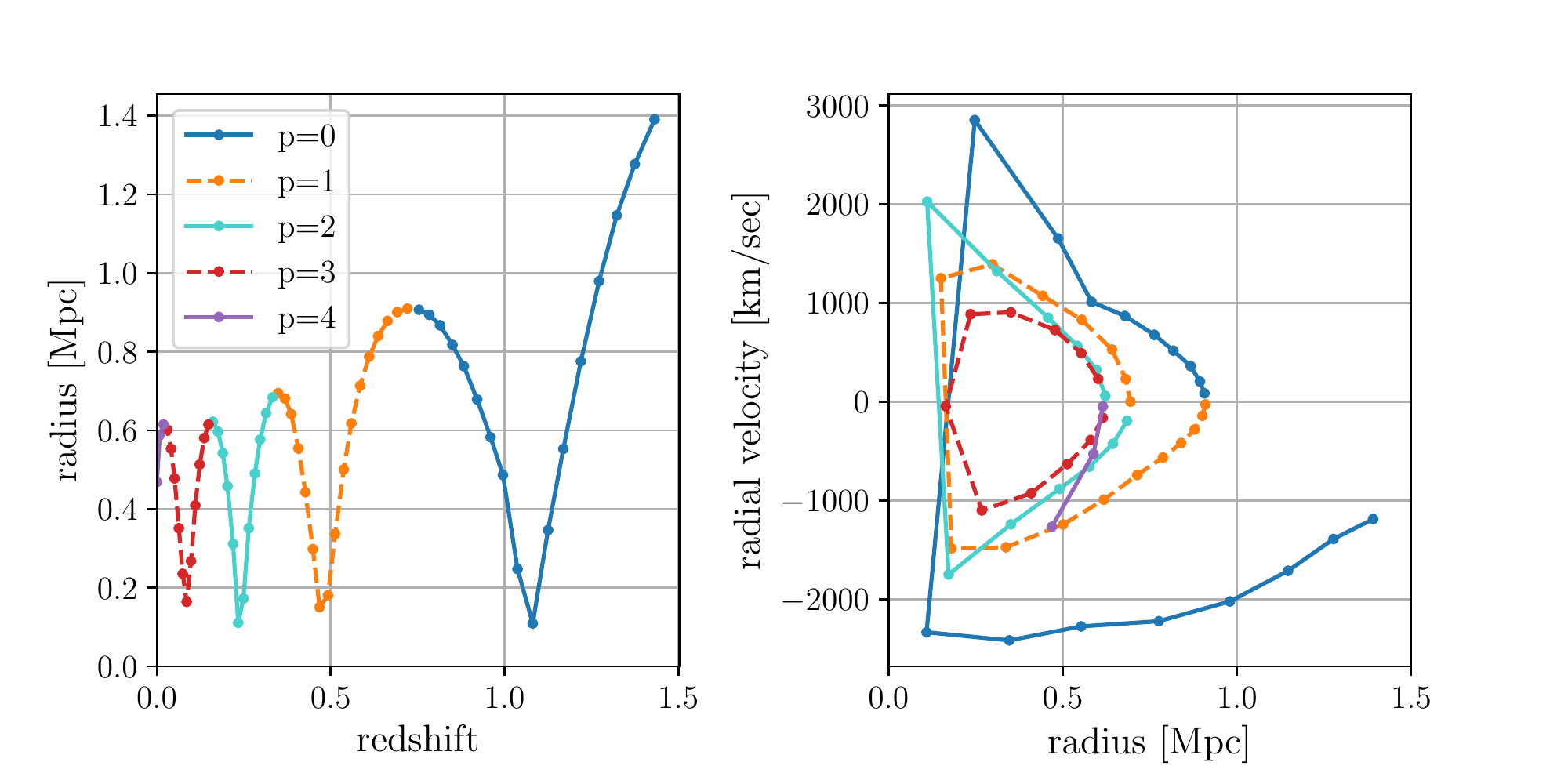}
  \caption{An example of particle trajectory for DM infalling into halo. Points with lines show the time evolution of a DM trajectory stored in the $N$-body snapshots. Left panel  plots the radial position as function of redshift, while right panel shows the trajectory in radial phase space.  Note that the colors indicate the different number of apocenter passages: $p=0$ (blue), $1$ (orange), $2$ (cyan), $3$ (red), and $4$ (purple).}
  \label{fig: particle_orbit}
\end{figure*}
%%%%%%%%%%%%%%%%%%%%%%%%%%%%%%%%%%%%%%%%%%%%%%%%%%%%%%%%%%%%%%%%%%%%%%%%%%%%%%%%%
%%%%%%%%%%%%%%%%%%%%%%%%%%%%%%%%%%%%%%%%%%%%%%%%%%%%%%%%%%%%%%%%%%%%%%%%%%%%%%%%%

%%%%%%%%%%%%%%%%%%%%%%%%%%%%%%%%%%%%%%%%%%%%%%%%%%%%%%%%%
%%%%%%%%%%%%%%%%%%%%%%%%%%%%%%%%%%%%%%%%%%%%%%%%%%%%%%%%%
\section{Method} 
\label{sec:method}
%%%%%%%%%%%%%%%%%%%%%%%%%%%%%%%%%%%%%%%%%%%%%%%%%%%%%%%%%
%%%%%%%%%%%%%%%%%%%%%%%%%%%%%%%%%%%%%%%%%%%%%%%%%%%%%%%%%

%%%--%%%--%%%--%%%--%%%--%%%--%%%--%%%--%%%--%%%--%%%--%%
%%%--%%%--%%%--%%%--%%%--%%%--%%%--%%%--%%%--%%%--%%%--%%
\subsection{$N$-body simulation}
\label{subsec:nbody}
%%%--%%%--%%%--%%%--%%%--%%%--%%%--%%%--%%%--%%%--%%%--%%
%%%--%%%--%%%--%%%--%%%--%%%--%%%--%%%--%%%--%%%--%%%--%%

\begin{table}
    \centering
    \begin{tabular}{c|c|c|c|c|c|c|c}
        $\Omega_m$ & $\Omega_r$ & $\Omega_b$ & $\sigma_8$ & $n_s$ & $h$ \\
       \hline
         0.99992   & 0.00008    & 0.04356    & 0.801 & 0.963      & 0.72 \\
\end{tabular}
    \caption{Summary of the cosmological parameters used in this paper. $\Omega_m$ is the matter density, $\Omega_r$ is the radiation density, $\Omega_b$ is the baryon density, $\sigma_8$ and $n_s$ give the normalisation and slope of the primordial matter power spectrum and $h$ is the hubble parameter.}
    \label{tb:cosmo_param}
\end{table}

We performed an $N$-body simulation with $N = ( 512 )^3$ particles distributed in a $(164.0625~h^{-1}\mathrm{Mpc})^3$ volume with an (almost) Einstein-de-Sitter cosmology. The choice of this cosmology is driven by the secondary infall model which is only valid in Einstein-de-Sitter cosmology. We however prefer to keep a radiation component so that our early universe calculation with the CAMB code \citep{lewis2000} remains accurate. This component is completely negligible in the late universe of interest in this paper ($z<2$). The cosmological parameters are given in Table~\ref{tb:cosmo_param}.

We use the same simulation set-up as in \citet{blot2014} focusing on a single realization. Initial conditions are generated at an initial redshift of $z_i=40$ with MPGRAFIC \citep{prunet2008} and assuming second-order Lagrangian Perturbation Theory (2LPT) for the displacement. The dynamical evolution of dark matter particles is calculated with RAMSES \citep{teyssier2002}. In order to trace the motions of particles, it is necessary to store enough snapshots of the simulation \citep{sparta1}. We stored 60 snapshots between redshifts 1.43 and 0, and labeled them by $n$ in ascending order of time. The snapshot are regularly spaced in expansion factor $a$ with $\delta a \approx 0.01$.  The snapshot $n = 40$ corresponds to $a=0.411$ or $z = 1.43$ and the snapshot $n = 99$ corresponds to $a=1$ or $z = 0$. 

From these snapshots we compute the density on a grid with $1024^3$ elements using a Cloud-In-Cell assignment scheme (CIC). The density at the location of the particles $\rho_i$ is linearly interpolated  from the density in the grid (i.e. using an inverse CIC scheme).  We detect halos at z=0 (only) with a parallel version (called pSOD) of the Spherical Overdensity (SO) halo finder algorithm \citep{lacey1994}. The center of halos is defined as the most-dense particle (which is close to the minimum of potential). A sphere is then grown around this center until the overdensity $\Delta_m=200$ (relative to the mean matter density in the universe) is reached. We found 11296 halos. After all halos are detected, we obtain a list of halo centers at z=0. Note that the size $R_{200}$ and mass $M_{200}$ of the SO halos as well as the location of SO halos at higher redshift do not play a role in the tracking procedure described below: this procedure only depends on the location of the center and the orbits of the particles around the center. This is in contrast from other tracking procedures (such as in \citet{sparta1}) where the tracking can start only after halo finders have been run on all snapshots.

%%%--%%%--%%%--%%%--%%%--%%%--%%%--%%%--%%%--%%%--%%%--%%
%%%--%%%--%%%--%%%--%%%--%%%--%%%--%%%--%%%--%%%--%%%--%%
\subsection{Tracking halos and particles}
\label{subsec:tracing_DM_particles}
%%%--%%%--%%%--%%%--%%%--%%%--%%%--%%%--%%%--%%%--%%%--%%
%%%--%%%--%%%--%%%--%%%--%%%--%%%--%%%--%%%--%%%--%%%--%%

In order to study the radial phase-space structure for each halo, we analyze snapshots densely sampled in time to keep track of the trajectories of dark matter particles. We, in particular, classify the trajectory of each dark matter particle by the number of apocenter passages experienced before $z=0$. To do so, we first need to identify the center of each halo at each snapshot, and then define the distance to each dark matter particle from the halo center, as well as the velocity of dark matter subtracting the bulk motion of the halo at the center. 
In Sec~.\ref{subsec: halo center},
we present the prescription to determine the location of the halo center at each snapshot. Then, in Sec.~\ref{sec: apocenter radius}, we analyze the particle trajectories with the velocity and position re-defined with respect to the halo centers.

%%%--%%%--%%%--%%%--%%%--%%%--%%%--%%%--%%%--%%%--%%%--%%
\subsubsection{Tracking of halo center}
\label{subsec: halo center}
%%%--%%%--%%%--%%%--%%%--%%%--%%%--%%%--%%%--%%%--%%%--%%

CDM halos typically have asymmetric shape with many substructures, and in a strict sense, the center of halo is not a well-defined notion. Nevertheless, we may identify the center-of-mass position near the most significant high-density region as a proxy of the halo center, and use it to keep track of the bulk motion of a halo. This would provide a robust estimate of the central part of a halo as long as we consider relatively massive halos.

We start with the halos identified at $z=0$ using SO algorithm. We track the identities of the particles near the center of mass back in time. The exact procedure is summarized as follows:
\begin{enumerate}
\item First, at z=0 data ($n=99$), pick up the $N_\mathrm{pickup}$ particles closest to the center position of halo. 
\item Go to one snapshot backward ($n=98$), and use the $N_\mathrm{pickup}$ particles identified previously to estimate their density-weighted center-of-mass position given below:
\begin{align}
    \mathbfit{x}_\mathrm{halo} = \sum_{i=1}^{N_\mathrm{pickup}} \frac{ \rho_i \mathbfit{x}_i }{ \rho_i } ,
\end{align}
where $\mathbfit{x}_i$ is the position of $i$-th DM particle, and $\rho_i$ is the local density at the particle.
\item Near the newly estimated center-of-mass position, pick up again the $N_\mathrm{pickup}$ closest particles at $n=98$.
\item Go to $n=97$ data, and use the $N_\mathrm{pickup}$ particles identified at $n=98$ to estimate their center-of-mass position.
\item Repeat the above steps until we reach the snapshot at $z=1.43$ ($n=40$).
\end{enumerate}

In this paper, we choose $N_\mathrm{pickup} = 128$ particles. The reason why we adopt the  density-weighted center-of-mass position is that rather than a true center-of-mass position, we wanted to know the densest region of the halo, which would be more stable against the merger event and any disturbances. We have checked that a robust estimation of the halo center is possible with the density-weighted method above, and the location of the halo center changes monotonically with time. 

After identifying the halo center at the snapshots $n=40-99$, the bulk velocity of the halo, $\mathbfit{v}_\mathrm{halo}$, is computed using these positions by the second-order finite difference method.

%%%--%%%--%%%--%%%--%%%--%%%--%%%--%%%--%%%--%%%--%%%--%%
\subsubsection{Identifying particle's apocenter passages}
\label{sec: apocenter radius}
%%%--%%%--%%%--%%%--%%%--%%%--%%%--%%%--%%%--%%%--%%%--%%

Having determined the halo center, we next focus on the trajectories of dark matter particles, and characterize their orbital motion with respect to the halo center, subtracting its bulk motion. In particular, we wish to clarify the multi-stream nature of CDM in phase space.

For this purpose, using the multiple snapshots, we identify the apocenter, and count the number of apocenter passages for each particle.  To do this,  we implement the SPARTA algorithm proposed by \citet{sparta1}. To be precise, this algorithm is originally used only to identify the splashback radius, i.e., the radius of the first apocenter passage. In this paper, we generalize the algorithm and apply it to identify the subsequent apocenter passages in the inner regions. 
That is, using the $60$ snapshots from $z=1.43$ to $0$, we keep track of each particle trajectory, and measure the radial velocity, $v_{\rm r}$, that is the difference in the peculiar velocities of the DM particle and the halo center of mass projected along the line of their separation. Namely, at the $n$-th snapshot, this is expressed as 
%%%%%%%%%%%%%%%%%%%%%%%%%%%%%%%%%%%%%%%%%%%%%%%%%%%%%%%%%%%%%%%%%%%%%%%%%%%%%%%%%
\begin{align}
v_{{\rm r},n}\equiv(\mbox{\boldmath$v$}_n-\mbox{\boldmath$v$}_{{\rm halo},n})\cdot\hat{r}_n,  
\end{align}
%%%%%%%%%%%%%%%%%%%%%%%%%%%%%%%%%%%%%%%%%%%%%%%%%%%%%%%%%%%%%%%%%%%%%%%%%%%%%%%%%
where $\mbox{\boldmath$v$}_n$, $\mbox{\boldmath$v$}_{{\rm halo},n}$, and $\hat{r}_n$ are the velocity of a DM particle, that of the halo center, and the unit vector pointing the DM particle from the halo center, respectively. The sign convention is such that the radial velocity defined above has a negative value for a particle approaching the halo center. The sign flips to positive when a particle passes the pericenter of the orbit. Conversely, a sign flip from positive to negative happens at the apocenter passage. The location of the first apocenter passage is particularly used to define the splashback radius \citep{sparta1}. We further keep tracking the sign flips of radial velocity beyond the first apocenter passage. Counting the number of apocenter passages $p$ for each particle, we classify the particle distribution in phase space by $p$, which is indeed useful to characterize the multi-stream structure of halos.

Fig.~\ref{fig: particle_orbit} shows an example of a particle trajectory extracted from our simulation. Based on the procedure mentioned above, the apocenter passages are identified, and the number of apocenter passages $p$ is incremented after passing through an apocenter (indicated in different colors). As shown in this figure, the procedure works well for isolated halos with a stationary accretion flow. However, when DM particles are captured by another halo or substructures, they may orbit around the center of this secondary gravitational source, not the center of the most prominent halos of our interest, relative to which the apocenter-passages should be examined. In such situations, the sign flip in the radial velocity can also occur due to the internal motion, not at the time of apocenter or pericenter passages. To avoid  misidentification of an apocenter passage, we thus monitor the direction of the relative position vector, $\hat{r}_n$, and require an additional condition that the vector must rotate by more than $\pi/2$ between adjacent apocenter passages. We checked that this ensures in most of the cases that the number $p$ is incremented only at the apocenter passage.

\begin{figure*}
    \centering
    \includegraphics[clip, width=138truemm, ]{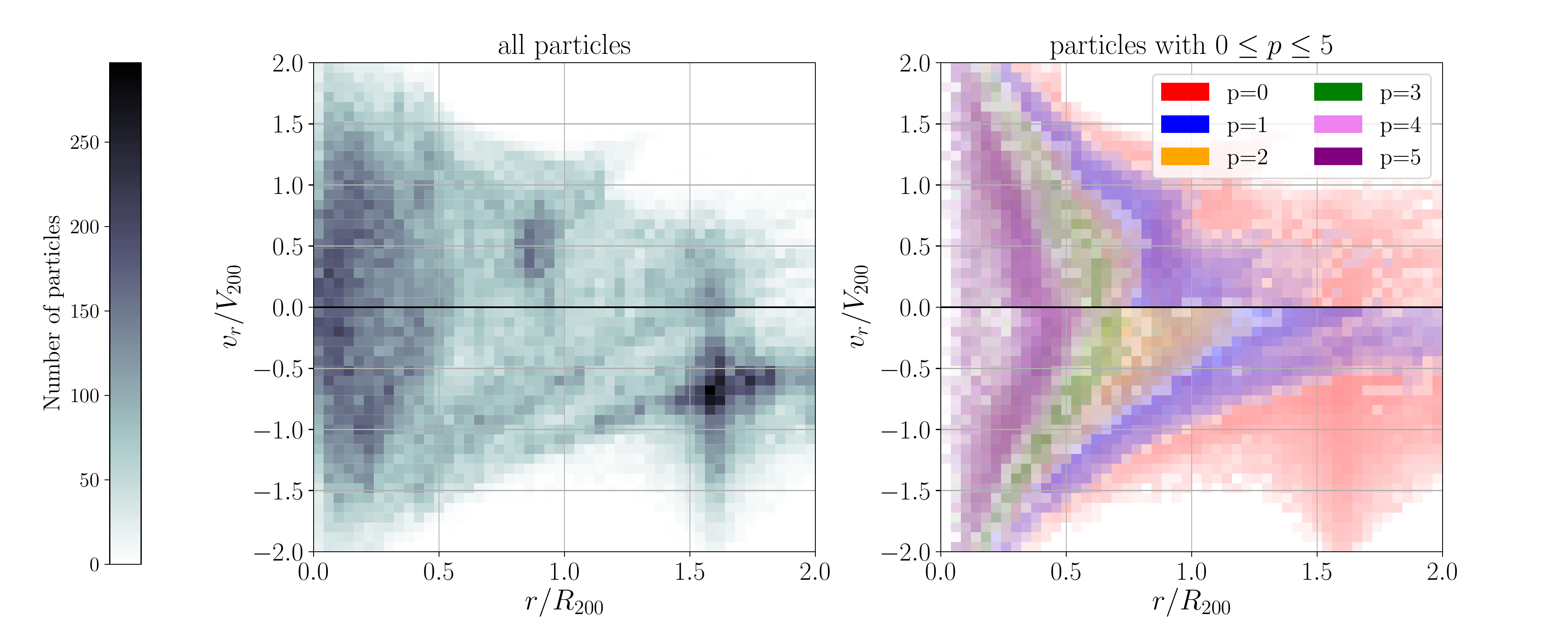}
    \includegraphics[clip, width=138truemm, ]{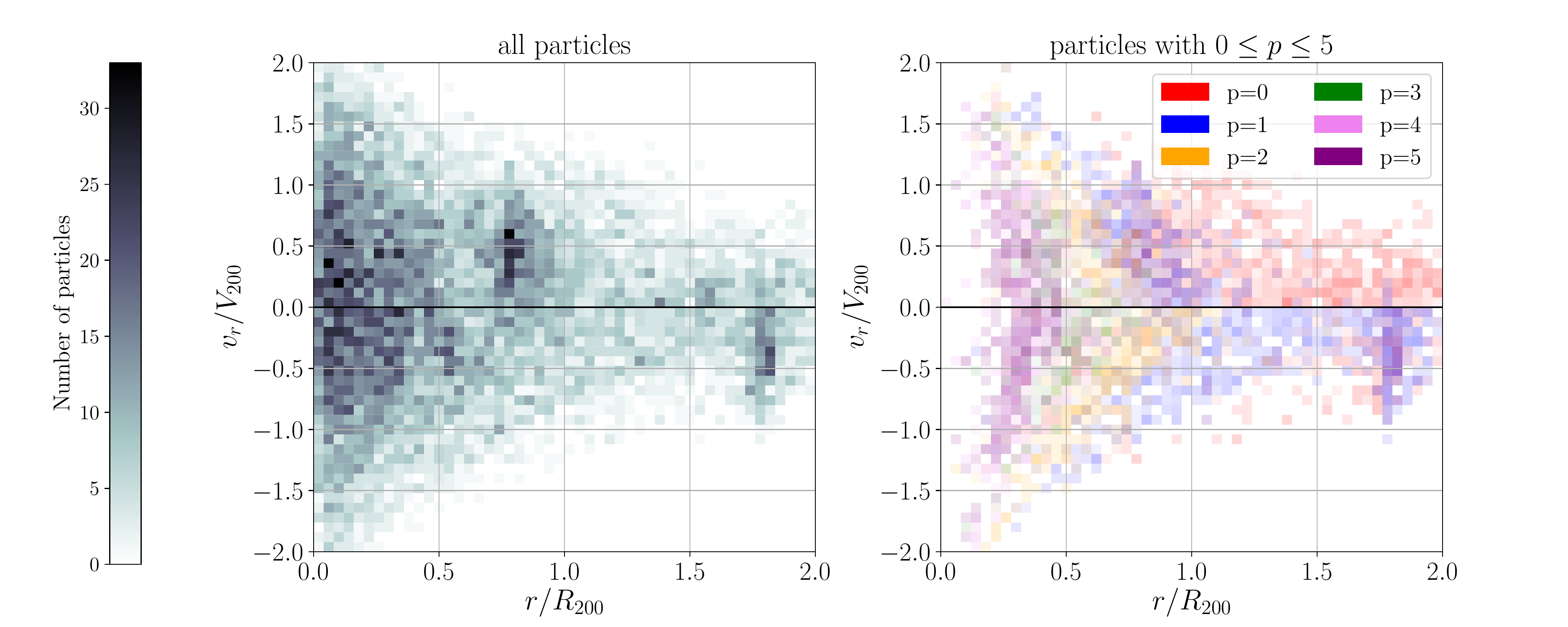}
    \includegraphics[clip, width=138truemm, ]{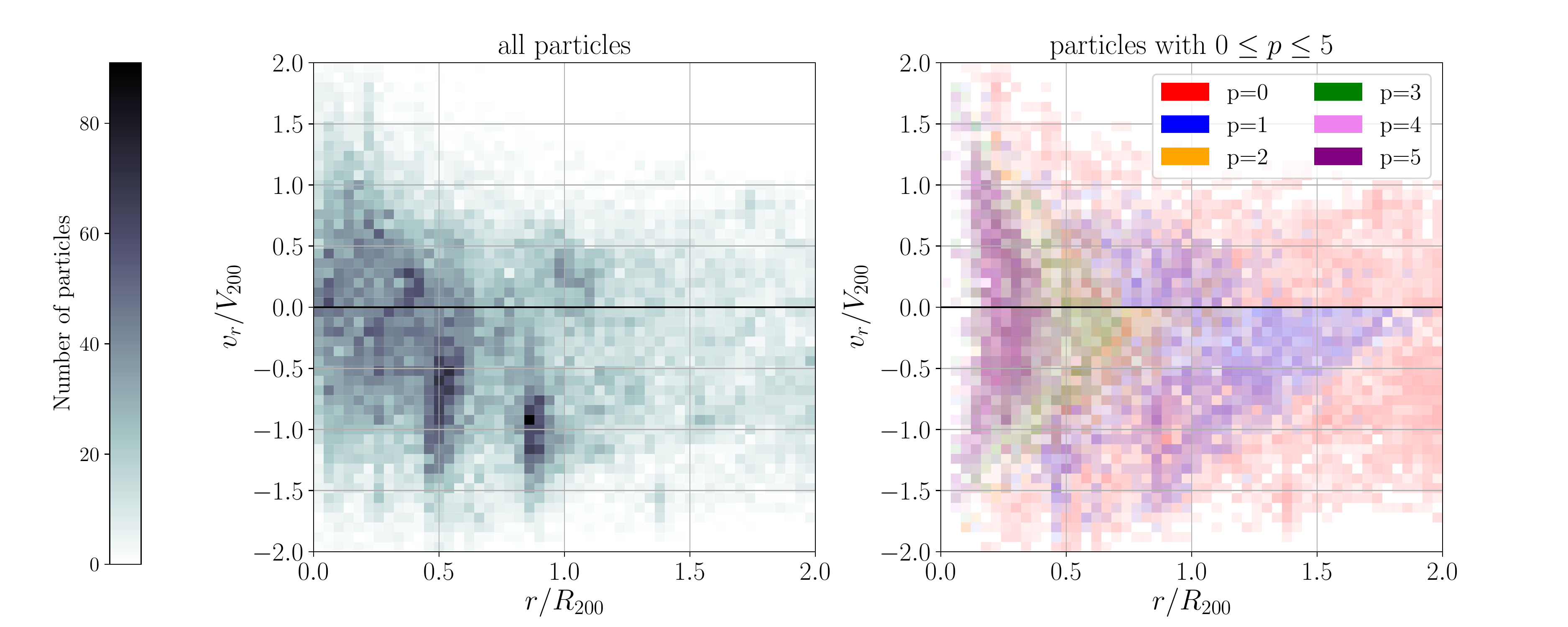}
    \includegraphics[clip, width=138truemm, ]{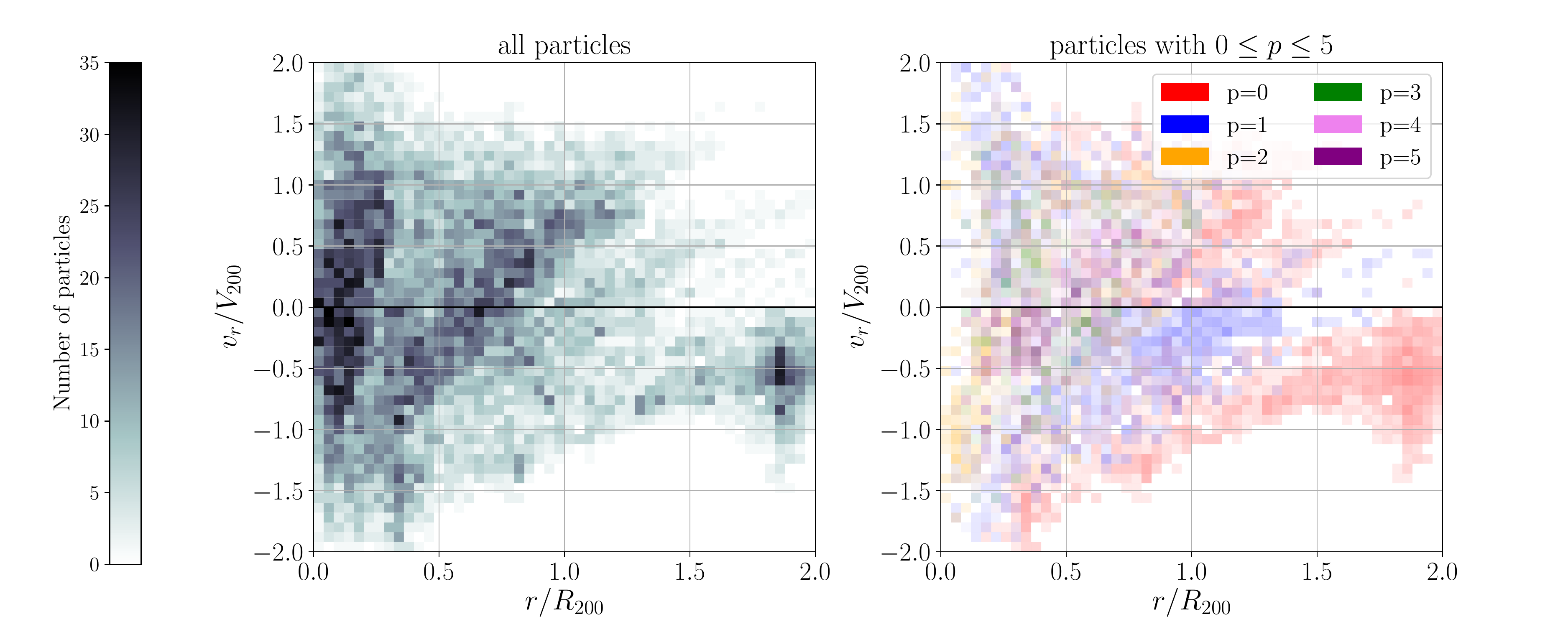}
    \caption{Radial phase-space distribution of DM particles for representative four halos. Left panels show the phase-space distibution for all DM particles near the selected halos without classification. Darker color indicates higher density. Right panels also plot the same phase-space distribution as shown in the left panels, but DM particles are classified with the number of apocenter passages, $p$, and are plotted in different colors. Note that in right panels, we plot only the particles with $p \leq 5$, and others with $p \geq6$ are removed.}
    \label{fig:phase-space-p-decomposition}
\end{figure*}

Finally, we repeat the procedure for all of the DM particles within $4 R_{200}$ at $z = 0$, and create, for each halo, the list of the number of apocenter passages $p$ for the DM particles, which is used to classify the particles in the phase-space distribution. Fig.~\ref{fig:phase-space-p-decomposition} shows the representative examples. Here, we select four specific halos, and plot for each halo the radial phase-space structure. Left panels show all DM particles near the halo, while in right panels, DM particles are classified with the number of apocenter passages, $p$, and plot them in different colors.

Comparing between left and right panels, we see that the bulk of the phase-space distribution is dominated by the particles with $p=0$, which are not properly the members of halo. The distribution of these DM particles exhibit irregular and extended structures in the presence of the merging halos/subhalos. On the other hand, except the last case (bottom panels), phase-space distributions of the particles with $p\geq 1$ look rather regular shape with a clear segregation of the particles with different $p$. Apart from the thick width of their distributions, each of the phase-space structures resembles the multi-stream features predicted by self-similar solution as shown in Fig.~\ref{fig: particle_orbit}.

To see more clearly, in Figs.~\ref{fig:halo_00019}-\ref{fig:halo_00841}, we separately plot in top panels the radial phase-space distributions tagged with the number of apocenter passages, $p$. Here, darker color implies higher density. Also, projected particle distributions in position space are shown in middle panels, while in bottom panels, the cumulative contribution of the density profile from the particles larger than $p$ is shown in different colors. As increasing $p$, we see clearly that particle distribution tends to get more clustered and rounder, though asymmetric features or substructures are also observed unlike the spherical self-similar solution. These trends motivate us to compare with the self-similar solution in more quantitative way. We will thus discuss in detail how to compare the measured phase-space distributions with self-similar solution in next subsection.

%%%--%%%--%%%--%%%--%%%--%%%--%%%--%%%--%%%--%%%--%%%--%%
%%%--%%%--%%%--%%%--%%%--%%%--%%%--%%%--%%%--%%%--%%%--%%
\subsection{Fitting the self-similar solution to the phase-space diagram}
\label{subsec: fitting method}
%%%--%%%--%%%--%%%--%%%--%%%--%%%--%%%--%%%--%%%--%%%--%%
%%%--%%%--%%%--%%%--%%%--%%%--%%%--%%%--%%%--%%%--%%%--%%

\begin{figure*}
  \centering
  \includegraphics[clip,height=58truemm]{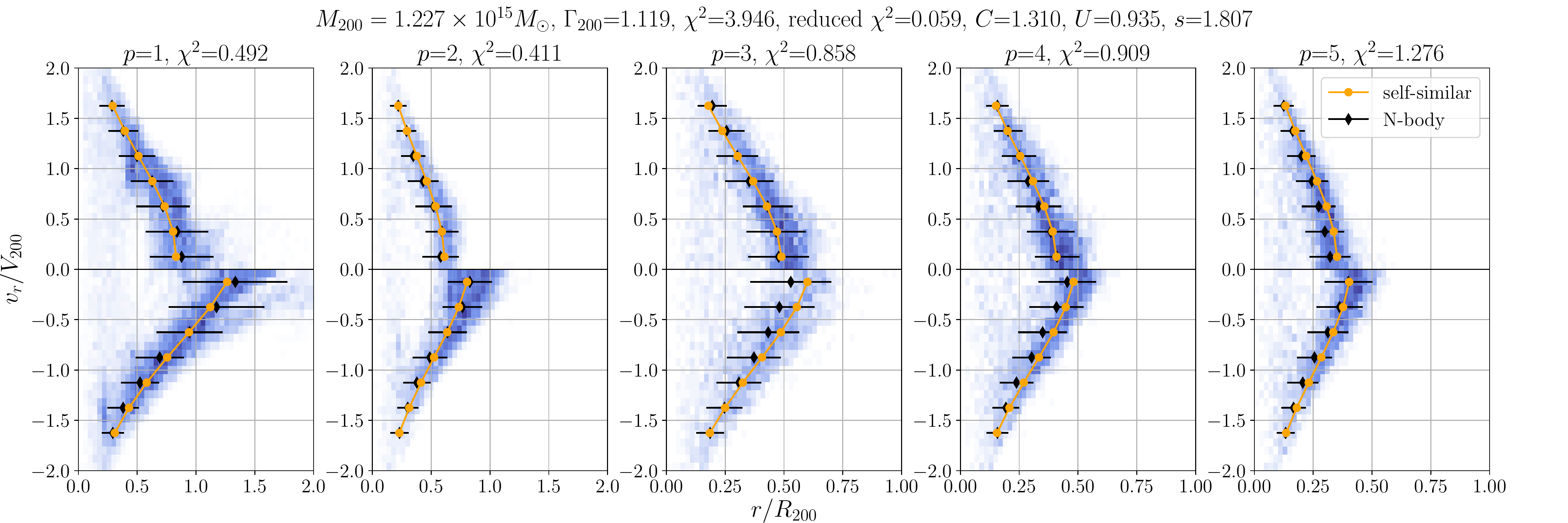}
  \includegraphics[clip,width=170truemm]{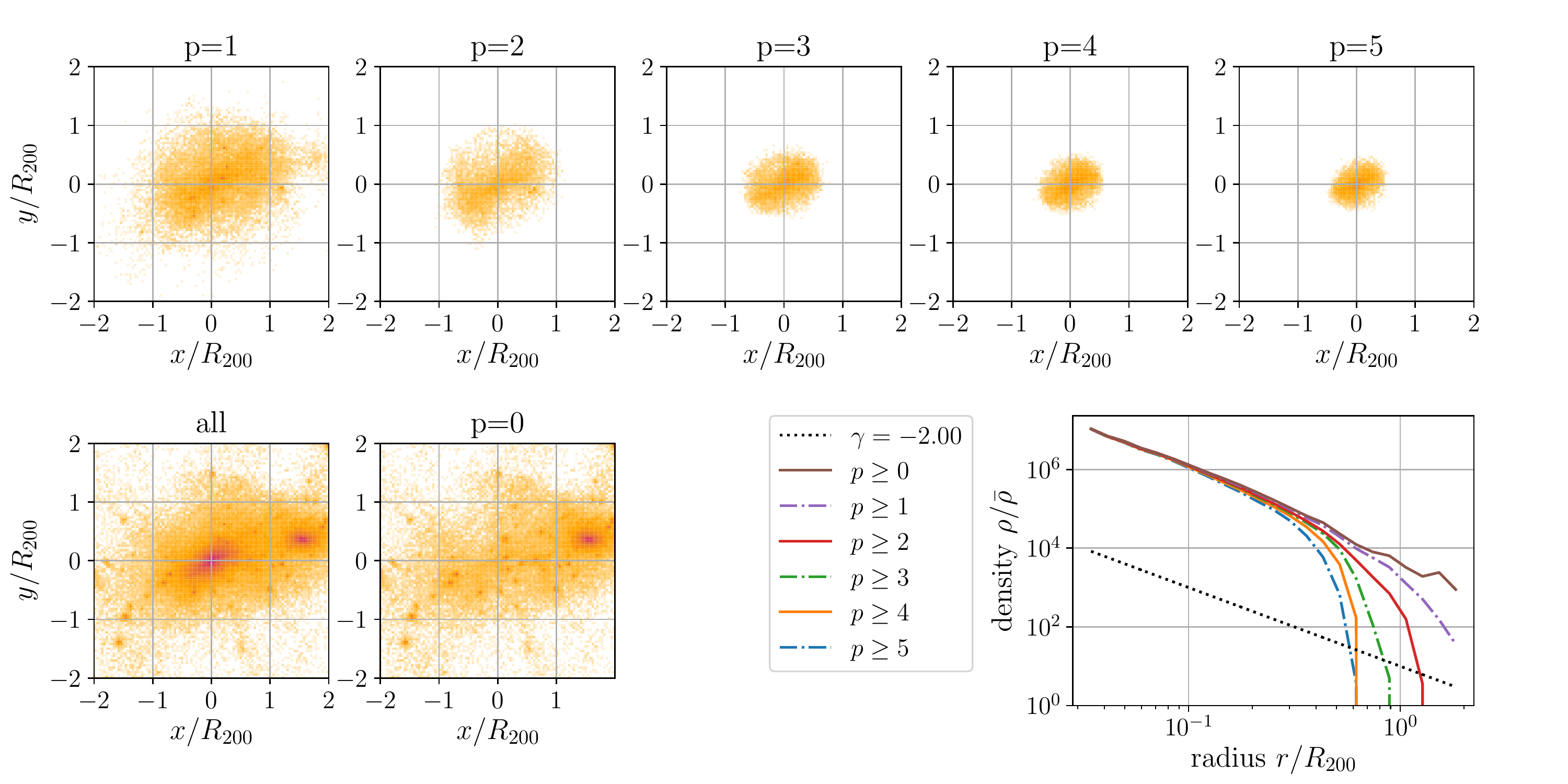}
  \caption{A cluster-size halo with $M_{200} = 1.227 \times 10^{15} M_\odot$,  which apparently shows a good agreement with self-similar solution (see Sec.~\ref{subsec:sample_selection}). This is the same halo as shown in the top panels of Fig.~\ref{fig:phase-space-p-decomposition}. 
    \textit{Top}: radial phase-space distribution of $N$-body particles with $p = 1, 2, \cdots, 5$ (denoted by blue colour contrast) with the best-fit self-similar solution (denoted by orange lines). Particles with different number of apocenter passages, $p$, are shown in different panels. Filled diamonds indicate the medians of $N$-body distributions in each velocity bin with error bars defined at Eq.~(\ref{error-bar}).
    \textit{Middle and bottom}: projected distribution of DM particles in position space (middle five panels and bottom two panels) and cumulative contribution to the radial density profile (bottom right panel), classified with number of apocenter passages, $p$.}
  \label{fig:halo_00019}
\end{figure*}

\begin{figure*}
  \centering
  \includegraphics[clip,height=58truemm]{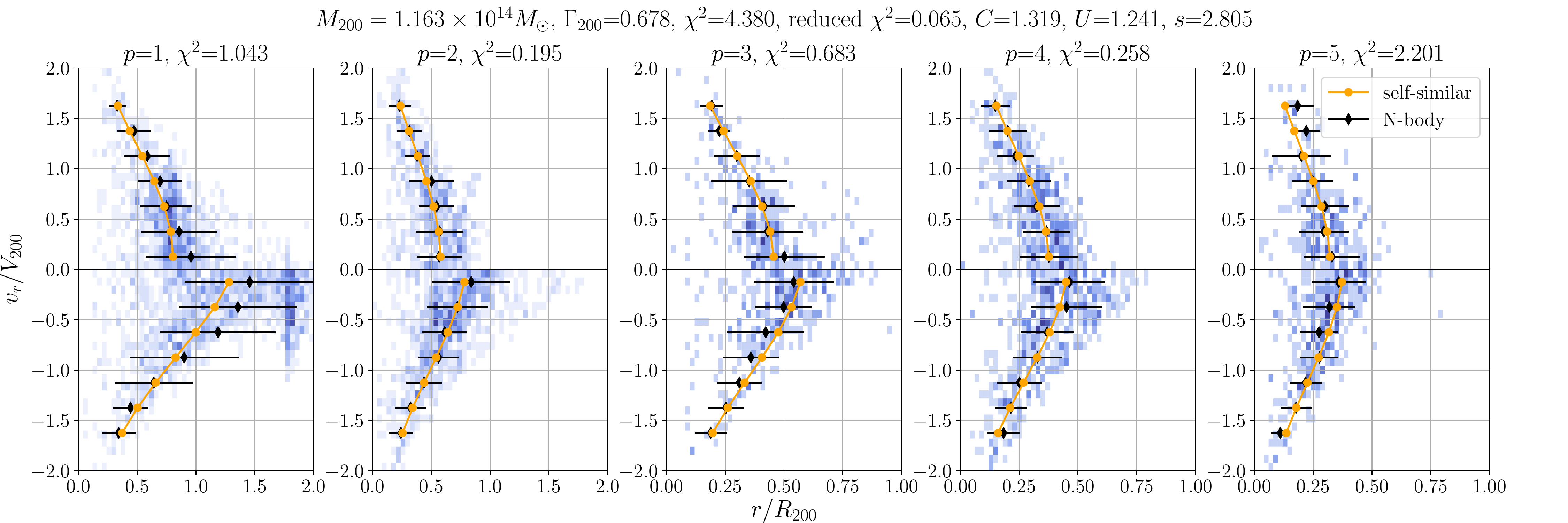}
  \includegraphics[clip,width=170truemm]{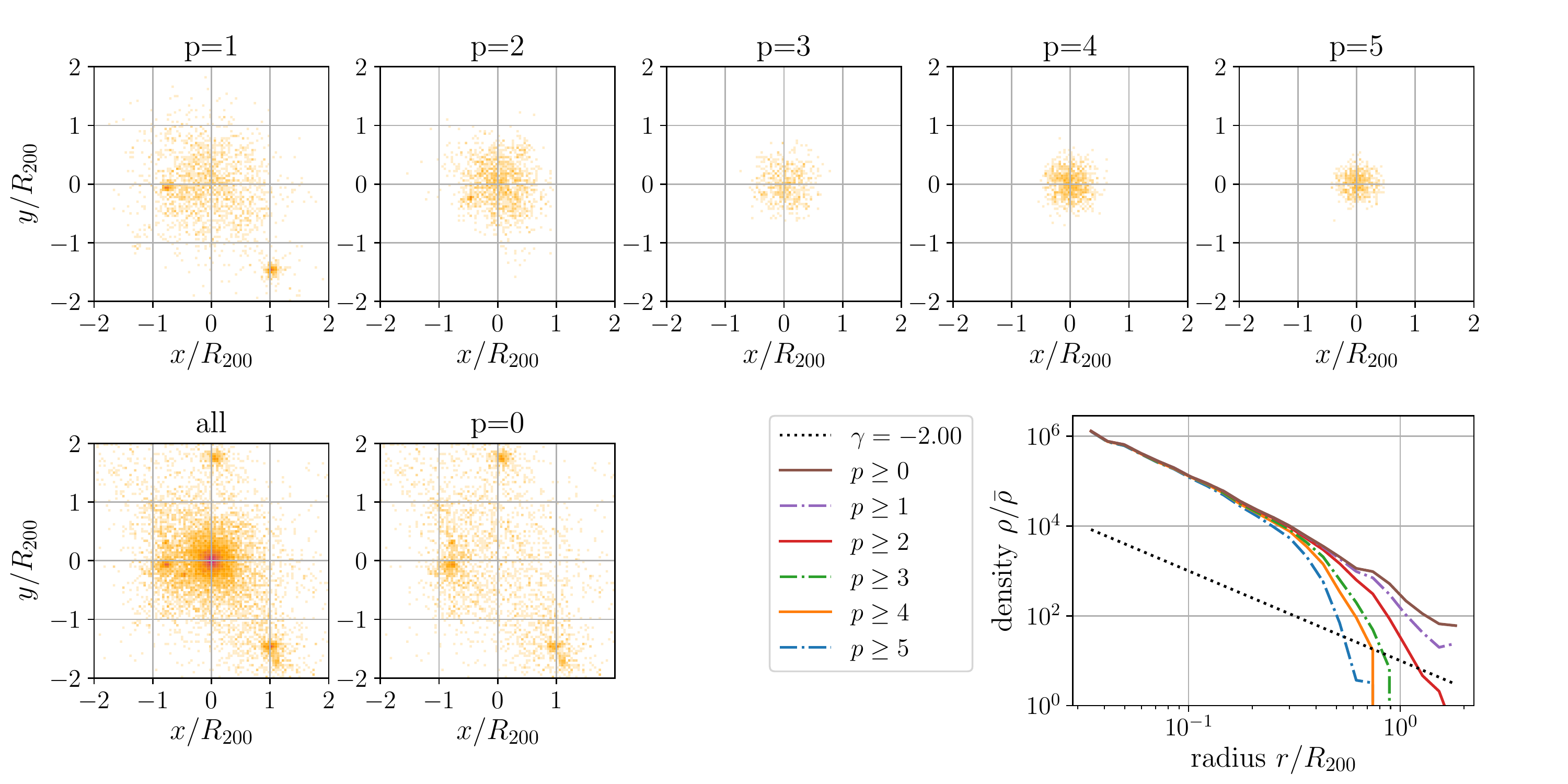}
  \caption{Same as in Fig.~\ref{fig:halo_00019}, but for a slightly less massive halo with  $M_{200}=1.163\times10^{14}\,M_\odot$. This is the same halo as shown in the second from the top in Fig~\ref{fig:phase-space-p-decomposition}.}
  \label{fig:halo_01244}
\end{figure*}

\begin{figure*}
  \centering
  \includegraphics[clip,height=58truemm]{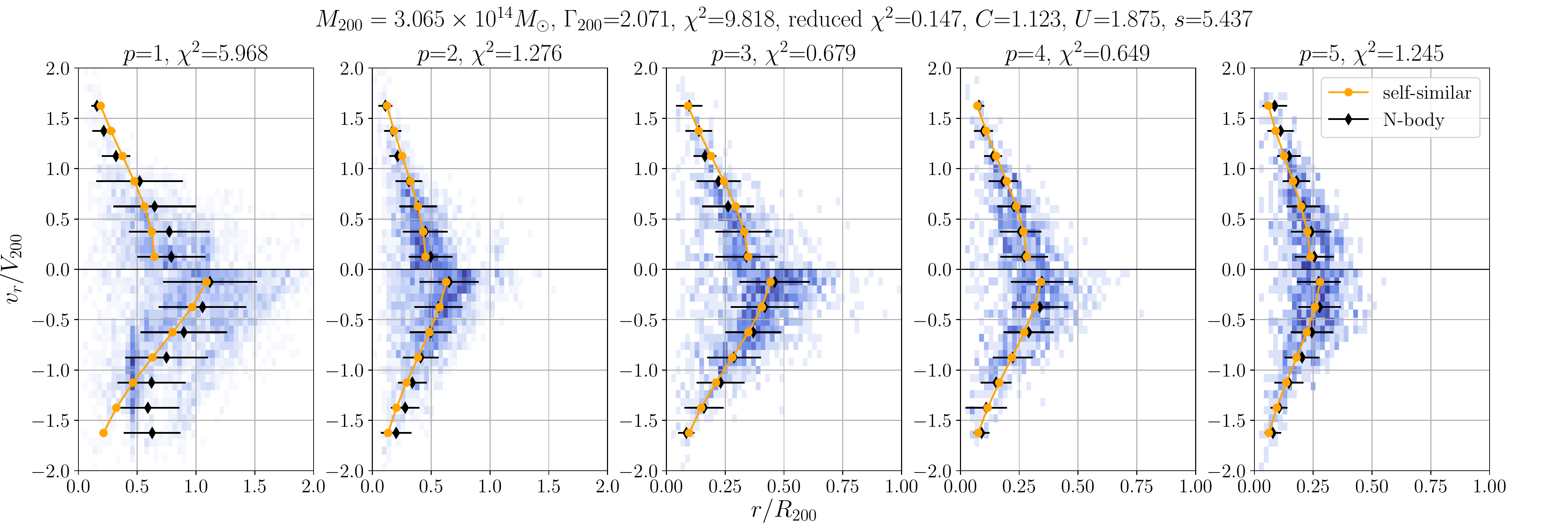}
  \includegraphics[clip,width=170truemm]{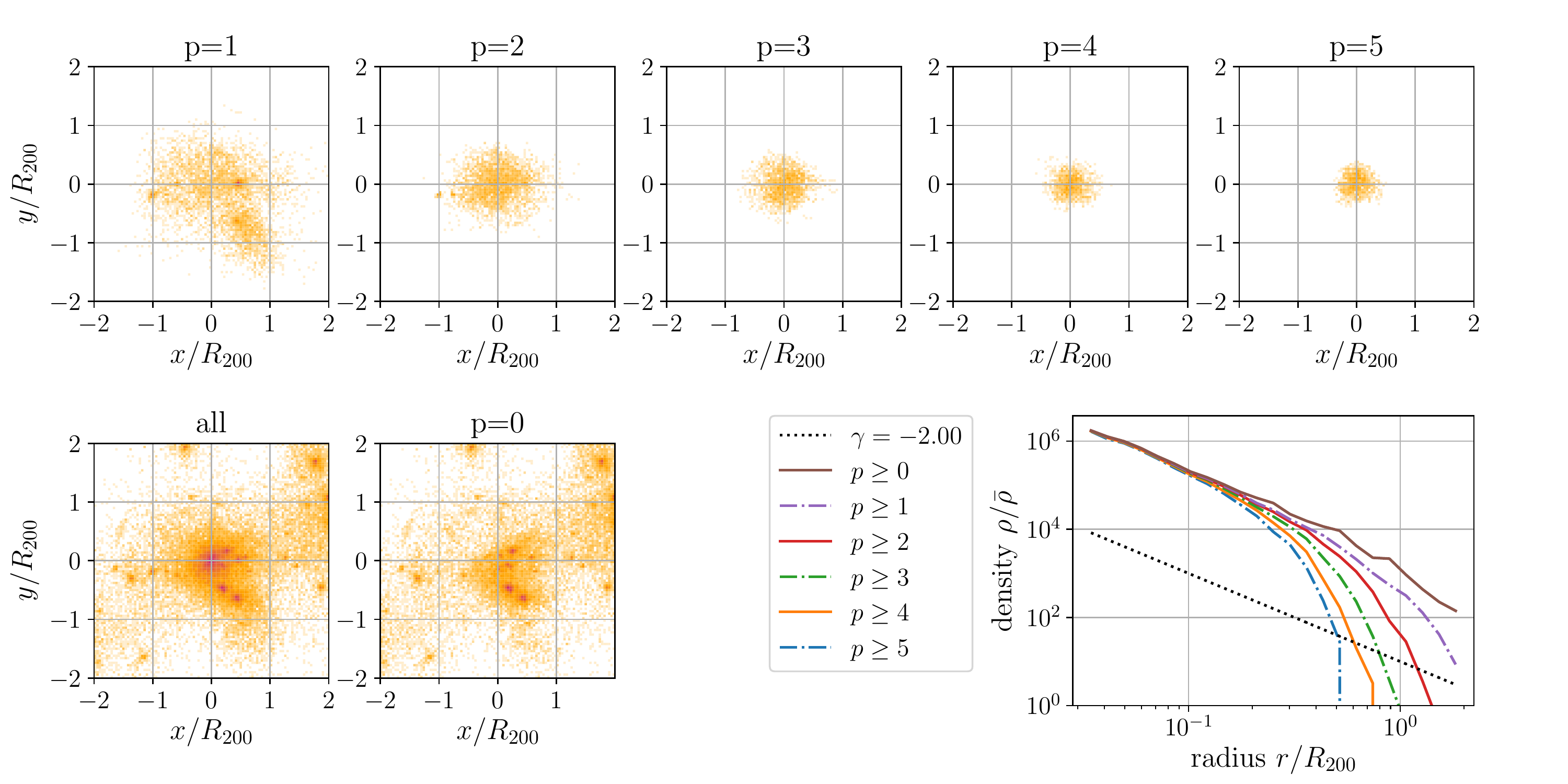}
  \caption{Same as in Fig.~\ref{fig:halo_00019}, but for a halo
    excluded by the condition (iii) given by Eq.~(\ref{chi2_p}) (See Sec.~\ref{subsec:sample_selection}). Note that this is the same halo as shown in the second from the bottom in Fig~\ref{fig:phase-space-p-decomposition}.}
  \label{fig:halo_00330}
\end{figure*}

\begin{figure*}
  \centering
  \includegraphics[clip,height=58truemm]{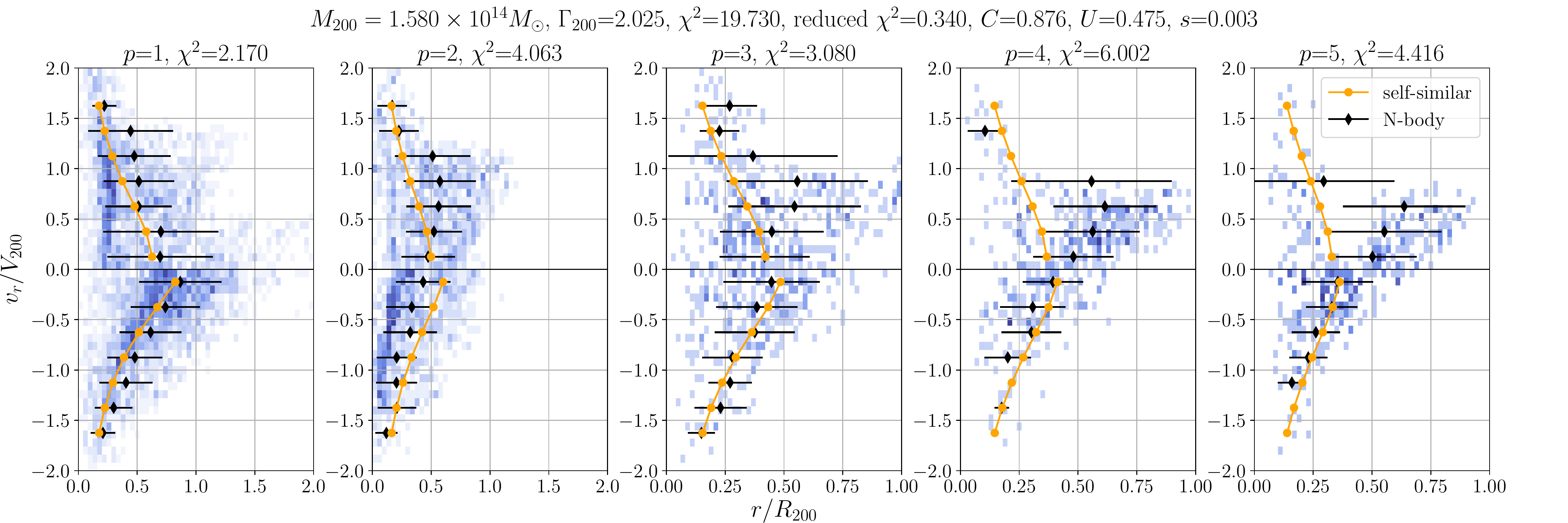}
  \includegraphics[clip,width=170truemm]{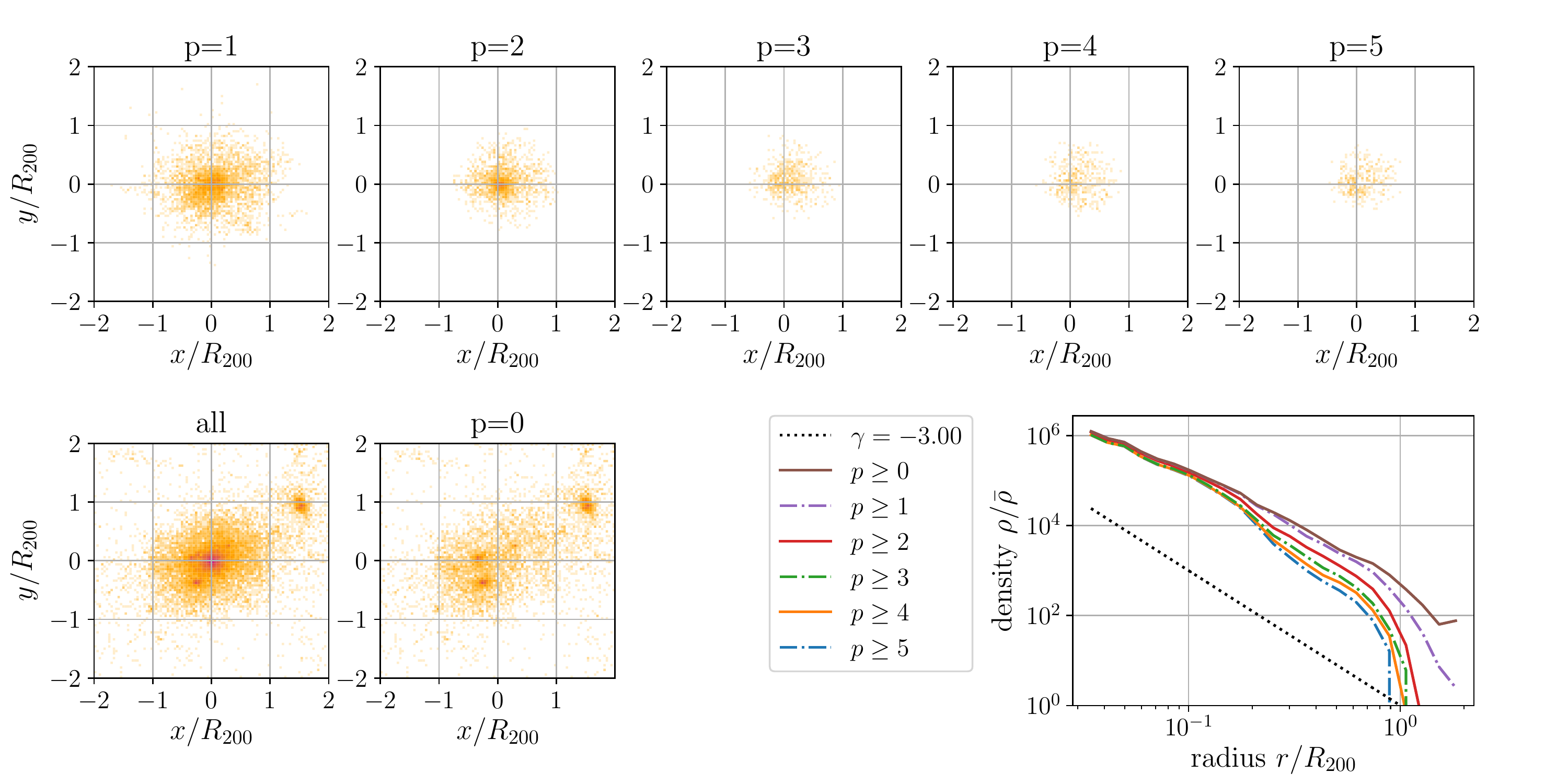}
  \caption{Same as in Fig.~\ref{fig:halo_00019}, but for a halo 
    excluded by the condition (ii). This is the same halo as shown in the bottom panel of Fig~\ref{fig:phase-space-p-decomposition}.}
  \label{fig:halo_00841}
\end{figure*}

We here describe the procedure to compare the simulation data with self-similar solution by \citet{fg84}. The self-similar solution provides both the time evolution of each mass element and the resultant snapshot of particle distribution in phase space at a given epoch [see Eq.~(\ref{phase-distribution})]. As shown in Figs.~\ref{fig:halo_00019}-\ref{fig:halo_00841}, we are particularly interested in characterizing the multi-stream nature of DM velocity flow, constructed with particle distributions tagged with the number of apocenter passages, $p$. Since the particles having the same value of $p$ are supposed to reside at the same stream line, we can conversely use the information on the apocenter passages for each DM particle to detect and identify the stream lines, whose location and shape can be predicted by the self-similar solution for a given set of model parameters. We shall thus fit the self-similar solution to the multiple stream lines for each halo in radial phase-space.

To best reproduce the multi-stream flow from self-similar solution, for each $p$, we divide the particle distribution in phase space into $14$ linearly-equal bins in radial velocity, ranging from $-(7/4) V_{200}$ to $(7/4) V_{200}$, where $V_{200} = \sqrt{ G M_{200} / R_{200} }$ is the circular velocity at the radius $R_{200}$.  The corresponding bin width is  $V_{200}/4$. For each velocity bin labeled by $i$, we use particles inside the bin to compute the median $r_{p, i}$ and the standard deviation $\sigma_{p, i}$ of the radial position. In top panels of Figs.~\ref{fig:halo_00019}-\ref{fig:halo_00841}, the estimated values of $r_{p,i}$ and $\sigma_{p,i}$ are depicted as filled black diamonds with errorbars. Large radial velocity bins tend to have small number of particles, which potentially lead to a biased estimation of median values. To compensate it, we inflate the error bars as
%%%%%%%%%%%%%%%%%%%%%%%%%%%%%%%%%%%%%%%%%%%%%%%%%%%%%%%%%%%%%%%%%%%%%%%%%%%%%%%%%
\begin{align}
  E_{p, i}^2 = \sigma_{p, i}^2 \left( 1 + \sqrt{ \frac{ 2 }{ n_{p, i} - 1 } } \right), 
  \label{error-bar}
\end{align}
%%%%%%%%%%%%%%%%%%%%%%%%%%%%%%%%%%%%%%%%%%%%%%%%%%%%%%%%%%%%%%%%%%%%%%%%%%%%%%%%%
where $n_{p, i}$ is the number of particles in the $i$-th radial velocity bin. Note that the second term at right-hand side of this equation is the ``error of error'' due to the Poisson noise. Since the fitting result is generally prone to be strongly affected by bins with small number of particles, the correction given above alleviates this to some extent. To be more conservative, we also ignore bins with $n_{p, i} < 5$, in fitting the data to self-similar solution.

Note that instead of the standard deviation given above, one may adopt the error on the mean in our fitting analyses given below. This would give us much smaller error bars by an extra $1/\sqrt{n_{p,i}}$ scaling, and one can test the spherical self-similar solution in a very strict sense. However, given the non-sphericity and the non-stationary accretion of halos in simulations, it is easy to expect that the $\chi^2$ values of the fitting using the error on the mean would be much larger than the number of degrees of freedom. We have confirmed this explicitly using some of the halos in our sample. In the same sense, the non-zero scatter in $r_{p,i}$ also implies that there exists no exact spherical halo with stationary accretion. Since we are rather interested in the bulk properties of each halo taking spherical average, we prefer to use the median of $r_{p,i}$ and the standard deviation at Eq.~(\ref{error-bar}) as the representative radial distance and spread in the particle distributions, and test the phase-space trajectories of DM particles in a statistical sense.

Having obtained the binned data set in radial velocity space for each $p$, we compare these data with self-similar solution expressed in the dimensionless coordinates as follows:
%%%%%%%%%%%%%%%%%%%%%%%%%%%%%%%%%%%%%%%%%%%%%%%%%%%%%%%%%%%%%%%%%%%%%%%%%%%%%%%%%
\begin{align}
  ( r/R_{200}, v_r/V_{200} ) = \left( C \Lambda(\tau), \ U \tau^{1-\beta} \frac{ d \lambda }{ d \tau } ( \tau ) \right),
    \label{trajectory with c and u}
\end{align}
%%%%%%%%%%%%%%%%%%%%%%%%%%%%%%%%%%%%%%%%%%%%%%%%%%%%%%%%%%%%%%%%%%%%%%%%%%%%%%%%%
with the function $\Lambda$ defined by $\Lambda(\tau) = \tau^{- \beta} \lambda ( \tau ) / \{\tau_\mathrm{sp}^{-\beta} \lambda ( \tau_\mathrm{sp} )\}$. Note that $\tau_\mathrm{sp}$ corresponds to the epoch of the first apocenter passage. Here, the quantities $C$ and $U$ are the scaling parameters for position and velocity, respectively. Comparison of  Eq.~(\ref{trajectory with c and u}) with Eq.~(\ref{phase-distribution}) implies $C =R_{\rm sp}/R_{200}$ and $U = \{r_\ta ( t )/t \}/V_{200}$, where $t$ is the age of the universe. In principle, the parameter $U$ can be determined once the values of $C$, $t$, $\beta$ or equivalently $s$, and $R_{200}$ are fixed. However, the relation between $U$ and other parameters assumes strict self-similarity and spherical symmetry during the entire history of halo evolution in an isolated setup. In particular, the age of the Universe $t$ in the self-similar solution corresponds to the age of halo counting from its formation time, which is somewhat ambiguous notion. Hence, in our fitting analysis, we do not relate $U$ with other parameters, but rather treat both $U$ and $C$ as independent free parameters.

To sum up, the free model parameters in self-similar solution are $C$, $U$ and $s$. These are determined by the likelihood analysis minimizing the function $\chi^2$: 
%%%%%%%%%%%%%%%%%%%%%%%%%%%%%%%%%%%%%%%%%%%%%%%%%%%%%%%%%%%%%%%%%%%%%%%%%%%%%%%%%
\begin{align}
  \chi^2 ( C, U, s )= \sum_{p = 1}^{p_\mathrm{max}} \sum_{i = 1}^{i_\mathrm{max}}
  \frac{ 1 }{ E^2_{p, i} } \left[ r_{p, i} - R_{200}\, \mathcal{R}_{i,p} (C,\, U,\, s) \right]^2 ,
  \label{chi2}
\end{align}
%%%%%%%%%%%%%%%%%%%%%%%%%%%%%%%%%%%%%%%%%%%%%%%%%%%%%%%%%%%%%%%%%%%%%%%%%%%%%%%%%
where $p$ and $i$ respectively run over the label of apocenter passages and the radial velocity bins, and we set $p_{\rm max}$ and $i_{\rm max}$ to $5$ and $14$, respectively. Note that the summation over the radial velocity bins in Eq.~(\ref{chi2}) is performed for the bins having more than five particles ($n_{p,i}\geq5$). 
Here, $r_{p,i}$ is the median value of the radial positions for particle data at $i$-th radial velocity bin with the number of apocenter passage $p$. The function $\mathcal{R}_{p,i}$ represents the prediction of self-similar solution, which is the radial position for the stream line corresponding to the number of apocenter passage $p$ at the $i$-th radial velocity bin, given by Eq.~(\ref{trajectory with c and u}).  For a given set of parameters, self-similar solution is computed, and the output results are tabulated numerically in the form of Eq.~(\ref{trajectory with c and u}) as function of $\tau$. Then, we can identify the stream line that corresponds to the $p$-th apocenter passage, from which we can further read off the radial position $r/R_{200}$ at the $i$-th radial velocity bin. In this way, we obtain  $\mathcal{R}_{i,p}$, which is finally plugged into Eq.~(\ref{chi2}). For an efficient computation of $\mathcal{R}_{i,p}$, we store the tabulated data set of self-similar solution finely sampled with every $0.1$ in parameter space of $s$, and linearly interpolate these data to obtain a new solution for the target value of $s$. We confirmed that the linearly interpolated results are converged to those obtained by quadratic interpolation and no significant difference is found.

Based on Eq.~(\ref{chi2}), we use the Markov-chain Monte Carlo (MCMC) algorithm to explore the model parameters for each halos, imposing the following uniform priors: 
%%%%%%%%%%%%%%%%%%%%%%%%%%%%%%%%%%%%%%%%%%%%%%%%%%%%%%%%%%%%%%%%%%%%%%%%%%%%%%%%%
\begin{align}
  C \in [ 0, 5 ] , \ \ U \in [ 0, 5 ], \ \ s \in [ 0, 9 ] . \label{mcmc-prior}
\end{align}
%%%%%%%%%%%%%%%%%%%%%%%%%%%%%%%%%%%%%%%%%%%%%%%%%%%%%%%%%%%%%%%%%%%%%%%%%%%%%%%%%
These parameter ranges are large enough that they do not affect the best-fit values and tails of the posterior distributions. Making use of the public python code, \textsc{emcee} \citep{emcee}, we calculated 4,000 steps with 12 walkers for all the 11,296 halos. The length of the chain would be sufficient to obtain convergent posterior distributions: the auto-correlation time of the MCMC chain is less than $1,000$ steps (typically a few hundred steps with slight variation among different halos).

\begin{figure}
  \includegraphics[clip,height=80truemm]{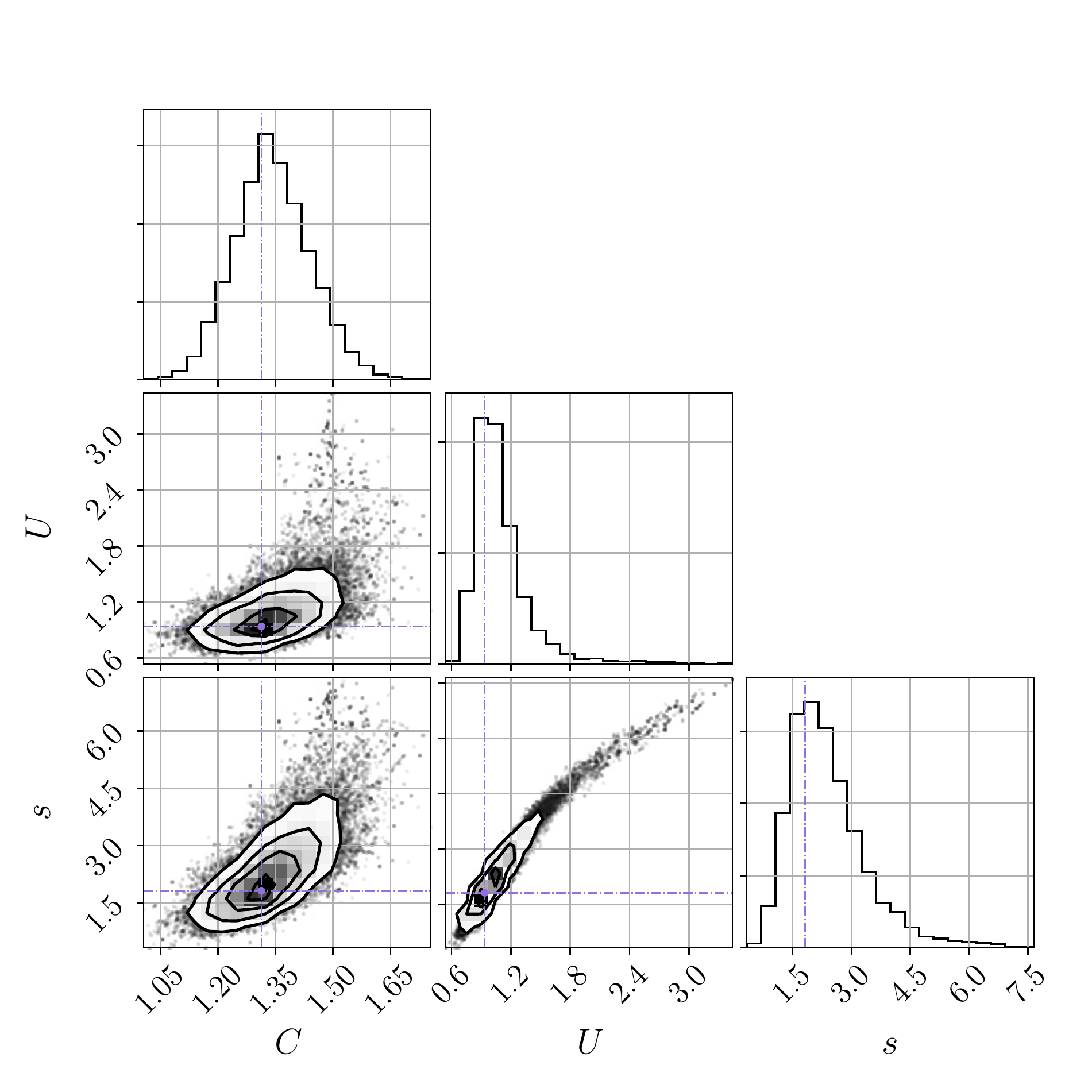}
  \caption{An example of the MCMC parameter estimation for a typical massive halo. The plot summarizes the marginalized two-dimensional posterior distribution for the parameters characterizing the self-similar solution, $C$, $U$ and the accretion rate $s$, discarding the steps in the burn-in period. Note that the parameters $C$ and $U$ are related to $R_{200}$ and $V_{200}$ through $C= R_{200} / R_{\rm sp}$ and $U= r_\mathrm{ta}/t/V_{200}$, where $R_{\rm sp}$ is the splashback radius, and $r_{\rm ta}$ is the turn-around radius. In each panel, the vertical and horizontal dot-dashed lines indicate the best-fit values of model parameters. Top panels summarize the one-dimensional projected posterior distribution for each parameter. Visualization of these MCMC results was made with Corner \citep{corner}.}
  \label{fig:halo_00019_mcmc_wop}
\end{figure}

For illustration, we show in Fig.~\ref{fig:halo_00019_mcmc_wop} the results of MCMC analysis for a cluster-size halo. The plotted results are the marginalized two-dimensional posterior distribution for the model parameters, 
discarding the first $800$ steps for each walkers as the burn-in period. As shown in Fig.~\ref{fig:halo_00019_mcmc_wop}, there is a unique maximum in density which is very close to the best-fit values of model parameters, depicted as the crossing point of the dot-dashed lines. We checked that the example shown here is typical, and the best-fit value is close to the peak position of posterior distribution. 
%%%%%%%%%%%%%%%%%%%%%%%%%%%%%%%%%%%%%%%%%%%%%%%%%%%%%%%%%%%%%%%%%%%%%%%%%%%%%%%%%
%%%%%%%%%%%%%%%%%%%%%%%%%%%%%%%%%%%%%%%%%%%%%%%%%%%%%%%%%%%%%%%%%%%%%%%%%%%%%%%%

%%%%%%%%%%%%%%%%%%%%%%%%%%%%%%%%%%%%%%%%%%%%%%%%%%%%%%%%%%%%%%%%%%%%%%%%%%%%%%%%%
%%%%%%%%%%%%%%%%%%%%%%%%%%%%%%%%%%%%%%%%%%%%%%%%%%%%%%%%%%%%%%%%%%%%%%%%%%%%%%%%%
%%%%%%%%%%%%%%%%%%%%%%%%%%%%%%%%%%%%%%%%%%%%%%%%%%%%%%%%%%%%%%%%%%%%%%%%%%%%%%%%%
%%%%%%%%%%%%%%%%%%%%%%%%%%%%%%%%%%%%%%%%%%%%%%%%%%%%%%%%%%%%%%%%%%%%%%%%%%%%%%%%%
%%%%%%%%%%%%%%%%%%%%%%%%%%%%%%%%%%%%%%%%%%%%%%%%%%%%%%%%%%%%%%%%%%%%%%%%%%%%%%%%%
%%%%%%%%%%%%%%%%%%%%%%%%%%%%%%%%%%%%%%%%%%%%%%%%%%%%%%%%%%%%%%%%%%%%%%%%%%%%%%%%%
%%%%%%%%%%%%%%%%%%%%%%%%%%%%%%%%%%%%%%%%%%%%%%%%%%%%%%%%%%%%%%%%%%%%%%%%%%%%%%%%%
%%%%%%%%%%%%%%%%%%%%%%%%%%%%%%%%%%%%%%%%%%%%%%%%%%%%%%%%%%%%%%%%%%%%%%%%%%%%%%%%%
%%%%%%%%%%%%%%%%%%%%%%%%%%%%%%%%%%%%%%%%%%%%%%%%%%%%%%%%%%%%%%%%%%%%%%%%%%%%%%%%%
%%%%%%%%%%%%%%%%%%%%%%%%%%%%%%%%%%%%%%%%%%%%%%%%%%%%%%%%%%%%%%%%%%%%%%%%%%%%%%%%%

%%%%%%%%%%%%%%%%%%%%%%%%%%%%%%%%%%%%%%%%%%%%%%%%%%%%%%%%%
%%%%%%%%%%%%%%%%%%%%%%%%%%%%%%%%%%%%%%%%%%%%%%%%%%%%%%%%%
\section{Results} 
\label{sec:result}
%%%%%%%%%%%%%%%%%%%%%%%%%%%%%%%%%%%%%%%%%%%%%%%%%%%%%%%%%
%%%%%%%%%%%%%%%%%%%%%%%%%%%%%%%%%%%%%%%%%%%%%%%%%%%%%%%%%

This section presents the main results of this paper, and gives a detailed comparison of the multi-stream flow of DM particles in the $N$-body simulation with the prediction of the self-similar solution from the phase-space point-of-view. Sec.~\ref{subsec:MCMC_results} presents the MCMC analysis based on Sec.~\ref{subsec: fitting method}. Sec.~\ref{subsec:sample_selection} presents the properties of the MCMC results for all of the halos identified at $z=0$, and discusses the selection of halo samples better fitted to the self-similar solution. Then, Sec.~\ref{subsec:correlation} shows the statistical properties of the model parameters for the well-fitted halos.

%%%--%%%--%%%--%%%--%%%--%%%--%%%--%%%--%%%--%%%--%%%--%%
%%%--%%%--%%%--%%%--%%%--%%%--%%%--%%%--%%%--%%%--%%%--%%
\subsection{Comparison of representative halos with self-similar solution}
\label{subsec:MCMC_results}
%%%--%%%--%%%--%%%--%%%--%%%--%%%--%%%--%%%--%%%--%%%--%%
%%%--%%%--%%%--%%%--%%%--%%%--%%%--%%%--%%%--%%%--%%%--%%

Here, for illustrated purpose, we pick up four representative halos among the total of $11,296$, and in the upper panels of Figs.~\ref{fig:halo_00019}-\ref{fig:halo_00841}, the binned phase-space distribution of the DM particles labeled by the number of apocenter passage $p$ (black filled circles with errorbars) is compared with the best-fitting self-similar solution, depicted as the yellow solid lines. Also, as we have seen in Sec.~\ref{sec: apocenter radius}, the middle and bottom-left panels of Figs.~\ref{fig:halo_00019}-\ref{fig:halo_00841} present the particle distributions for each $p$ on a two-dimensional projected position space, while the bottom-right panels plot the density profiles for the cumulative contributions of the particles experienced at least $p$ apocenter passages. 

The first and second example of the halos, shown in Figs.~\ref{fig:halo_00019} and \ref{fig:halo_01244}, are the phase-space structure well fitted by the self-similar solution, with the best-fit values of $s$ being $s=1.81$ and $2.81$, respectively. The mass of these halos are $M_{\rm 200}=1.23\times 10^{15}\,M_\odot$ and $1.16\times 10^{14}\,M_\odot$, respectively.  As it is clear from the figures, the predictions with the best-fitting parameters reproduce the measured phase-space distribution binned along the velocity axis remarkably well. The particle distributions of these halos in position space, seen in the bottom-left panels of Figs.~\ref{fig:halo_00019} and \ref{fig:halo_01244}, exhibit substructures or clumps, and their global shape is indeed asymmetric. Nevertheless, as increasing $p$, the particle distributions gets smoother, and tend to be rounder. Further, the best-fit values of $s$ indicate that the asymptotic slope of the density profile is $-2$, i.e., $\rho\propto r^{-2}$ [see Eq.~(\ref{eq:asymptotic_slope})], which in fact agrees well with inner slope of the measured density profile, shown in the bottom right panels of Figs.~\ref{fig:halo_00019} and \ref{fig:halo_01244}.

The third example, shown in Fig.~\ref{fig:halo_00330}, is a halo with mass $3.07\times 10^{14}\,M_\odot$, three times larger than the second example. Although the best-fitting self-similar solution seems to explain the overall trends of the binned phase-space distribution from the simulation well, a closer look at the simulation data at $p=1$ reveals a structure elongated vertically at $v_{\rm r}/V_{\rm 200}<-1$, and a systematic discrepancy between the simulation and the model is manifest around this structure. This corresponds in position space to a large blob at $(x,y)\simeq(0.5\,R_{200},0)$ shown in the leftmost panel in the middle row. The discrepancy is mostly ascribed to this component with significant scatter in the radial velocity, just experienced the first apocenter passage after a major merger. Also, in the position plot for the particles with $p=0$ (second from the left in the bottom row), we can observe several significant substructures near the center. These are before the merger to the main halo as indicated by the fact that they have $p=0$. The existence of these features might also have disturbed the orbits of the already accreted DM particles. Since the self-similar solution by \citet{fg84} describes an isolated halo with stationary accreting matter, this is, in a sense, a typical example violating the basic assumption of the model. 

On the other hand, the fourth example, shown in Fig.~\ref{fig:halo_00841},  has a mass similar to the second example with a much smaller value of the best-fit parameter, $s\simeq0$. Visually, the agreement between the self-similar solution and the simulation is bad. In contrast to the third example, a large discrepancy is now found in the phase-space distribution at $p>3$. Because of this, the inner slope of the measured density profile does not agree well with that of the best-fit self-similar solution (see the bottom right panel). Looking at the particle distribution in position space at $z=0$, we find that unlike the previous examples, the spatial extent of the particle distribution does not shrink with increasing $p$ for $p\geq3$. Although we do not see any clear signature of the clumps or substructures at $p\geq1$, we suspect that the discrepancy is due to the remnant of orbiting substructures which is tidally stripped. In fact, going back to the snapshots at slightly earlier time, we confirm that this halo underwent a major merger with a small impact parameter, and the infalling halo exhibited a rapid orbital decay followed by the tidal stripping. Thus, the example shown here may not be regarded as a relaxed halo, though it is difficult to judge only from the spatial distribution at the final snapshot. In this respect, the phase-space distribution is more informative, and is powerful to probe the dynamical properties of halo structure.

In Figs.~\ref{fig:halo_00019}-\ref{fig:halo_00841}, we also show the values of both $\chi^2$ and the reduced $\chi^2$ for the best-fit model. Note that the number of degree of freedom to derive the reduced $\chi^2$ varies by halos because of the different number of available velocity bins, but it typically ranges from $50$ to $70$. In agreement with visual inspection of the radial phase-space plots, the first example has the smallest $\chi^2$ among the four representative examples, and the fourth example has the largest $\chi^2$ value. However, their reduced $\chi^2$ values are rather small, and both are less than $1$. These small values basically come from the rather loose error we adopted in estimating $\chi^2$ [see Eq.~(\ref{error-bar}) and the subsequent paragraph on our choice of the error bars]. Thus, in order to study the properties of halos ``well fitted'' by the self-similar solution in our sense, the $\chi^2$ values or the reduced $\chi^2$ values alone are insufficient. We have to come up with other additional requirements or criteria to form a sample of well-fitted halos. 

%%%--%%%--%%%--%%%--%%%--%%%--%%%--%%%--%%%--%%%--%%%--%%
%%%--%%%--%%%--%%%--%%%--%%%--%%%--%%%--%%%--%%%--%%%--%%
\subsection{Sample selection}
\label{subsec:sample_selection}
%%%--%%%--%%%--%%%--%%%--%%%--%%%--%%%--%%%--%%%--%%%--%%
%%%--%%%--%%%--%%%--%%%--%%%--%%%--%%%--%%%--%%%--%%%--%%

Applying the method described in Sec.~\ref{sec:method}, we have analyzed $11,296$ halos whose virial masses $M_{200}$ are greater than $10^{13} M_\odot$.  As we have seen in Sec.~\ref{subsec:MCMC_results}, the self-similar solution sometimes fails to describe the multi-stream structure of phase-space distribution for halos in $N$-body simulation. A part of the reason is ascribed to the fact that some halos near the low-mass end do not have  sufficient number of particles to determine the location of the streams to be compared in detail with the self-similar solutions. In order to quantitatively clarify the extent to which the self-similar solution can describe the multi-stream feature of halos in radial phase space, one may introduce strict selection criteria for each halo well fitted by the self-similar solution. Although this leaves us only a biased subset of simulated halos, their statistics would give us useful insight on the structure of more realistic halos.

First condition we impose is that the number of radial velocity bins having DM particles more than five should be at least $48$ out of $70$ over $p=1-5$ [condition (i)]. This excludes $2,924$ halos, leaving 8,372. Next, we exclude the halos for which the radial position of the stream line is not well-determined like those shown in Fig.~\ref{fig:halo_00841} ($p=3-5$). This can be originated from different reasons: a significant fraction of particles failed to be assigned the correct number of apocenter passages due to the limitation of our algorithm, or the actual phase space distribution is far from self-similar solutions due to the major merger, a large number of substructures, or highly asymmetric shape. The $\chi^2$ defined at Eq.~(\ref{chi2}) alone cannot perfectly isolate these halos as ``badly fitted'' because a poor determination of the particle trajectories generally leads to a large value of $E_{p,i}$ [see Eq.~(\ref{error-bar})]. We thus impose another condition to exclude those halos from the later analyses as follows. For each stream line and at each radial velocity bin, we compute the ratio, $E_{p, i}/\bar{r}_{p, i}$, where $\bar{r}_{p, i}$ is the median value of radial position for particles in the $i$-th velocity bin for particles after the $p$-th apocenter passage. This ratio indicates how well we can determine the median location of the stream line. We then exclude the halo in which the seventh largest value of this ratio is greater than 0.625 [condition (ii)]. With this condition, 4,108 halos are excluded.

Applying the conditions mentioned above, we now assess the {\it goodness of fit} 
using the minimum value of $\chi^2$ obtained from the MCMC analysis. We impose [condition (iii)]
%%%%%%%%%%%%%%%%%%%%%%%%%%%%%%%%%%%%%%%%%%%%%%%%%%%%%%%%%%%%%%%%%%%%%%%%%%%%%%%%%
\begin{align}
  [ \chi^2 ]_p \leq 3.5, \ \ (p = 1, \cdots, 5),
  \label{chi2_p}
\end{align}
%%%%%%%%%%%%%%%%%%%%%%%%%%%%%%%%%%%%%%%%%%%%%%%%%%%%%%%%%%%%%%%%%%%%%%%%%%%%%%%%%
where the subscript $p$ indicates that $\chi^2$ is computed only for the particles with $p$ apocenter passages. The halo shown in Fig.~ \ref{fig:halo_00330} is a typical example excluded by this third condition, and a significant deviation from the self-similar solution is found for the first apocenter passage $p=1$, $\chi^2_{p=1}=5.968$. Note that with this last condition, halos whose phase-space particle distribution apparently resembles the best-fitting self-similar solution are sometimes excluded. In this respect, the resultant samples that meet all the selection criteria may be regarded as conservative and high-quality halos well described by self-similar solution. We label these halos as ``well-fitted'' samples.

%%%%%%%%%%%%%%%%%%%%%%%%%%%%%%%%%%%%%%%%%%%%%%%%%%%%%%%%%%%%%%%%%%%%%%%%%%%%%%%%%
%%%%%%%%%%%%%%%%%%%%%%%%%%%%%%%%%%%%%%%%%%%%%%%%%%%%%%%%%%%%%%%%%%%%%%%%%%%%%%%%%
\begin{table}
  \centering
  \begin{tabular}{l|c}
   \hline
    Conditions & Number of halos \\ \hline \hline
    None & 11,296 (100\%)\\ \hline
    (i) Sufficient particles in most of the bins & 8,372 (74.1\%)\\ \hline
    (i) + (ii) Good orbit determination & 4,264 (37.0\%) \\ \hline
    (i) + (ii) + (iii) Well-fitted by the self-similar solution & 3,561 (31.5\%) \\ \hline
  \end{tabular}
  \caption{The number of halos meeting our selection conditions.}
  \label{tb:halo_selection}
\end{table}
%%%%%%%%%%%%%%%%%%%%%%%%%%%%%%%%%%%%%%%%%%%%%%%%%%%%%%%%%%%%%%%%%%%%%%%%%%%%%%%%%
%%%%%%%%%%%%%%%%%%%%%%%%%%%%%%%%%%%%%%%%%%%%%%%%%%%%%%%%%%%%%%%%%%%%%%%%%%%%%%%%%

%%%%%%%%%%%%%%%%%%%%%%%%%%%%%%%%%%%%%%%%%%%%%%%%%%%%%%%%%%%%%%%%%%%%%%%%%%%%%%%%%
%%%%%%%%%%%%%%%%%%%%%%%%%%%%%%%%%%%%%%%%%%%%%%%%%%%%%%%%%%%%%%%%%%%%%%%%%%%%%%%%%
\begin{figure*}
  \centering
  \includegraphics[clip,height=48truemm]{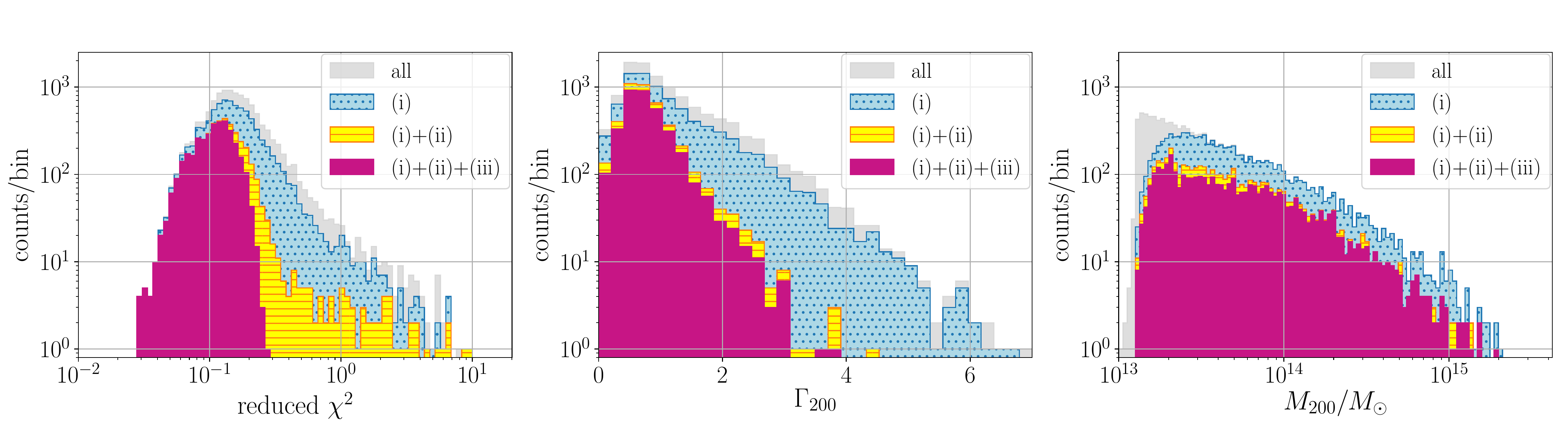}
  \caption{Frequency distributions of halos against reduced $\chi^2$ (left), accretion rate $\Gamma_{200}$ (middle), halo mass $M_{200}$ (right) 
    for halos which meet our selection criteria (see Table \ref{tb:halo_selection}). Vertical axis represents the number of halos per bin.}
  \label{fig:halo_properties}
\end{figure*}
%%%%%%%%%%%%%%%%%%%%%%%%%%%%%%%%%%%%%%%%%%%%%%%%%%%%%%%%%%%%%%%%%%%%%%%%%%%%%%%%%
%%%%%%%%%%%%%%%%%%%%%%%%%%%%%%%%%%%%%%%%%%%%%%%%%%%%%%%%%%%%%%%%%%%%%%%%%%%%%%%%%

Table \ref{tb:halo_selection} summarizes the number of halos that meet each of the selection criteria. 
To see how our criteria gives (un-)biased halo samples, we plot in Fig.~\ref{fig:halo_properties} the frequency distribution of halos against the quantities characterizing the individual halo properties. The left panel shows the distribution against the reduced $\chi^2$. We see that the condition (i) preferentially removes halos having a rather large value of reduced $\chi^2$. Combining the condition (ii) further excludes halos mainly with large reduced $\chi^2$, but there still remain halos with a moderately large reduced $\chi^2$ survived. Adding the third condition, those halos are finally removed, and the resultant frequency distribution exhibits a sharp cutoff around the reduced $\chi^2\sim0.3$, which is consistent with Eq.~(\ref{chi2_p}) for the individual orbit specified by $p$ given that the total degree of freedom over $1\leq p\leq 5$ in the likelihood analysis is roughly around $50-70$. 

The middle panel of Fig.~\ref{fig:halo_properties} shows the frequency distribution against the mass accretion rate, $\Gamma_{200}$, directly measured from the $N$-body simulation, which is defined as follows \citep[e.g.,][]{dk14}: 
%%%%%%%%%%%%%%%%%%%%%%%%%%%%%%%%%%%%%%%%%%%%%%%%%%%%%%%%%%%%%%%%%%%%%%%%%%%%%%%%%
\begin{align}
  \Gamma_{200} := \frac{ \Delta \ln M_{200} }{ \Delta \ln a } , \ \ \
  \Delta X \equiv X ( z = 0 ) - X ( z = 0.5 ) ,
  \label{eq:Gamma200}
\end{align}
%%%%%%%%%%%%%%%%%%%%%%%%%%%%%%%%%%%%%%%%%%%%%%%%%%%%%%%%%%%%%%%%%%%%%%%%%%%%%%%%%
The definition above has been used in the literature as an indicator to characterize the environmental dependence of the splashback radii on top of the rather trivial mass dependence. A notable feature seen in the frequency distribution is that the condition (ii), which rejects halos with large uncertainties in the locations of stream shells, almost determines the accessible range of $\Gamma_{200}$ for the final samples, excluding the halos having a large value of $\Gamma_{200}$. This implies that a rapid mass accretion tends to disturb the trajectories of DM particles inside the halo, thus leading to a wider stream line/shell, i.e., the radial distribution of particles having the same value of $p$ for a given radial velocity.

Finally, the right panel of Fig.~\ref{fig:halo_properties} shows the frequency distribution against the halo mass $M_{200}$. While the condition (i) removes light halos almost only in the range $M_{200}\lesssim 2\times 10^{13}\,M_\odot$, the other two conditions does not change the shape of the distribution. As a result, the final sample of halos can be regarded as a representative sample of the original in terms of mass, except for the lightest end. This is in marked contrast to the effect of the selection on the accretion rate.

To conclude, one should keep in mind that our final sample is biased toward low accretion rate, but nearly representative in terms of the halo mass in later analyses. Once these are in mind, a large number of halos that meet all conditions would allow us to study statistical properties of the multi-stream nature of CDM halos.

%%%--%%%--%%%--%%%--%%%--%%%--%%%--%%%--%%%--%%%--%%%--%%
%%%--%%%--%%%--%%%--%%%--%%%--%%%--%%%--%%%--%%%--%%%--%%
\subsection{Statistical properties of well-fitted halo samples}
\label{subsec:correlation}
%%%--%%%--%%%--%%%--%%%--%%%--%%%--%%%--%%%--%%%--%%%--%%
%%%--%%%--%%%--%%%--%%%--%%%--%%%--%%%--%%%--%%%--%%%--%%

%%%%%%%%%%%%%%%%%%%%%%%%%%%%%%%%%%%%%%%%%%%%%%%%%%%%%%%%%%%%%%%%%%%%%%%%%%%%%%%%%
%%%%%%%%%%%%%%%%%%%%%%%%%%%%%%%%%%%%%%%%%%%%%%%%%%%%%%%%%%%%%%%%%%%%%%%%%%%%%%%%%
\begin{figure}
  \centering
  \includegraphics[clip,width=80truemm]{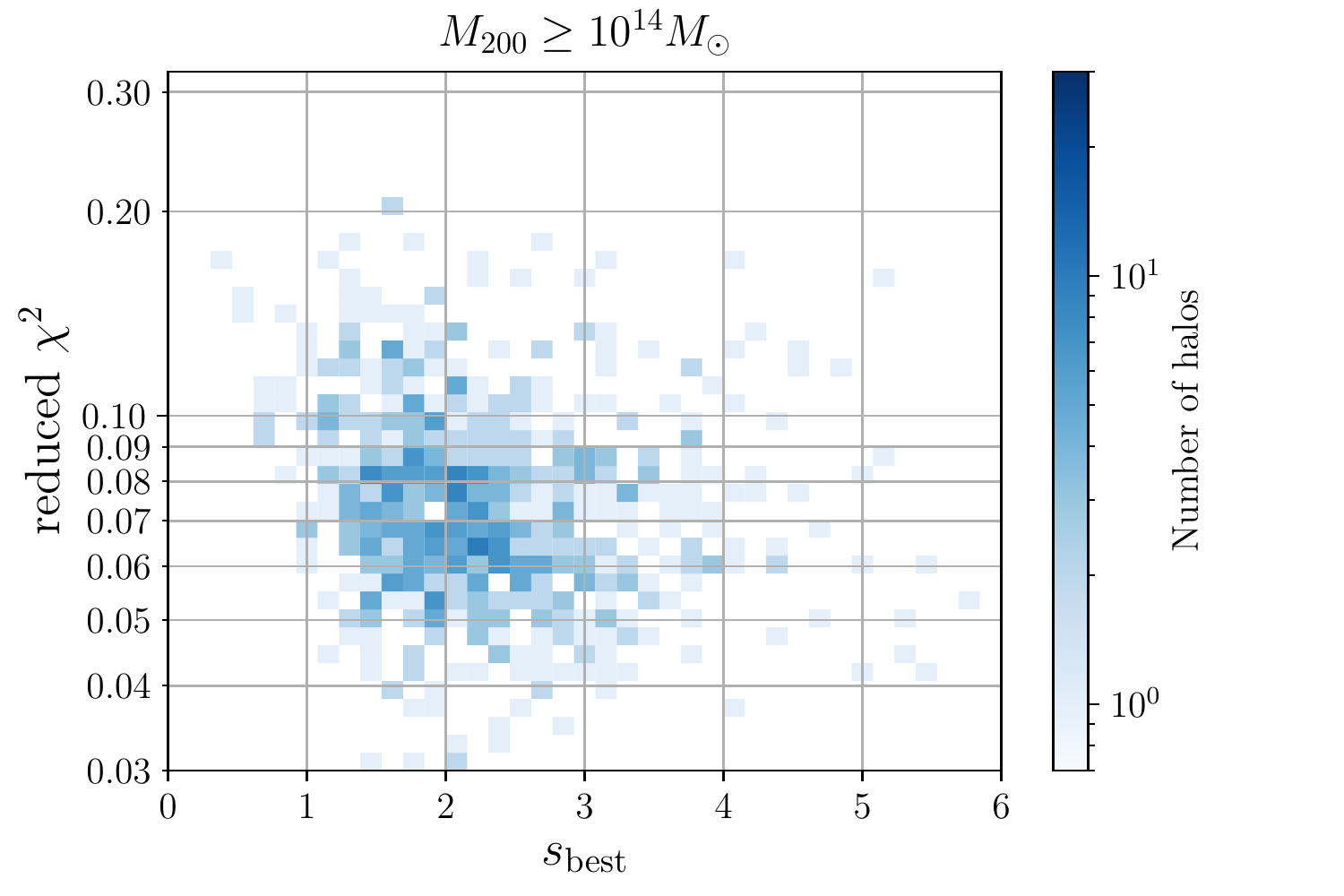}
  \includegraphics[clip,width=80truemm]{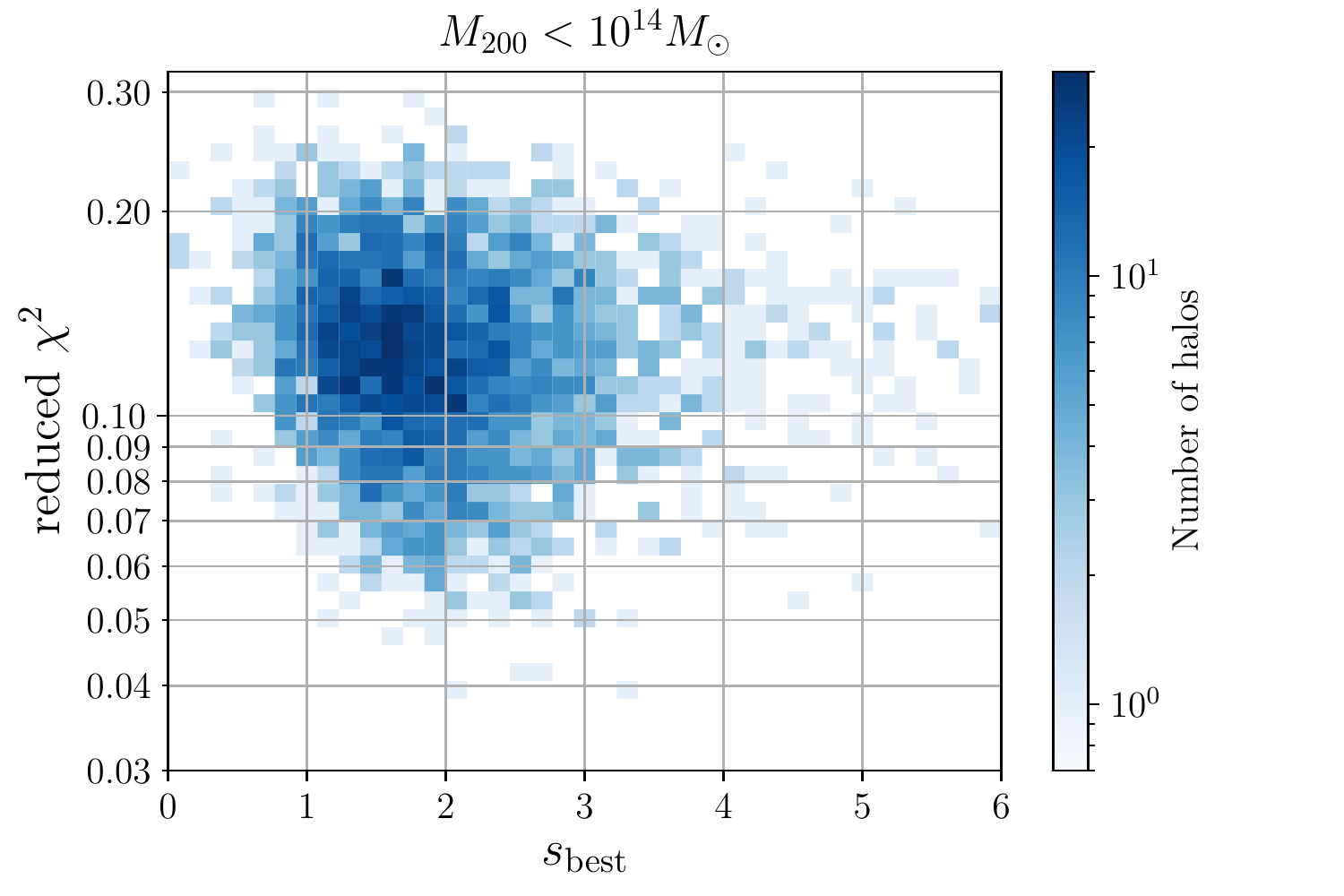}
  \caption{Frequency distribution of halos plotted in two-dimensional plane of the reduced $\chi^2$ and best-fit parameter of $s$, $s_{\rm best}$. Top and bottom panels show the results for well-fitted halo samples with mass greater and less than $10^{14}\,M_\odot$, respectively. Color depth in each pixel indicates the number of halos falling into the pixel in logarithmic scales.}
  \label{fig:xs}
\end{figure}
%%%%%%%%%%%%%%%%%%%%%%%%%%%%%%%%%%%%%%%%%%%%%%%%%%%%%%%%%%%%%%%%%%%%%%%%%%%%%%%%%
%%%%%%%%%%%%%%%%%%%%%%%%%%%%%%%%%%%%%%%%%%%%%%%%%%%%%%%%%%%%%%%%%%%%%%%%%%%%%%%%%

The halo sample selected in Sec.~\ref{subsec:sample_selection} are characterized not only by the measured values of the mass $M_{200}$ and accretion rate $\Gamma_{200}$ from usual SO halo definition, but also by the best-fit parameters in the self-similar solution, i.e., the accretion rate parameter $s_{\rm best}$, and the two dimensionless quantities $C_{\rm best}$ and $U_{\rm best}$. Note that $C_{\rm best}$ represents the ratio of the splashback radius to the radius $R_{200}$, i.e., $R_{\rm sp}/R_{200}$. As shown in Fig.~\ref{fig:halo_00019_mcmc_wop}, the parameter $U_\mathrm{best}$ is strongly correlated with $s_\mathrm{best}$. Hence, focusing on four other physical parameters, $M_{200}$, $\Gamma_{200}$, $s_\mathrm{best}$ and $C_\mathrm{best}$, and also the reduced $\chi^2$ of the best-fit model, we examine the statistical properties of the selected halos.

%%%%%%%%%%%%%%%%%%%%%%%%%%%%%%%%%%%%%%%%%%%%%%%%%%%%%%%%%%%%%%%%%%%%%%%%%%%%%%%%%
%%%%%%%%%%%%%%%%%%%%%%%%%%%%%%%%%%%%%%%%%%%%%%%%%%%%%%%%%%%%%%%%%%%%%%%%%%%%%%%%%
\begin{figure}
  \centering
  \includegraphics[clip,height=65truemm]{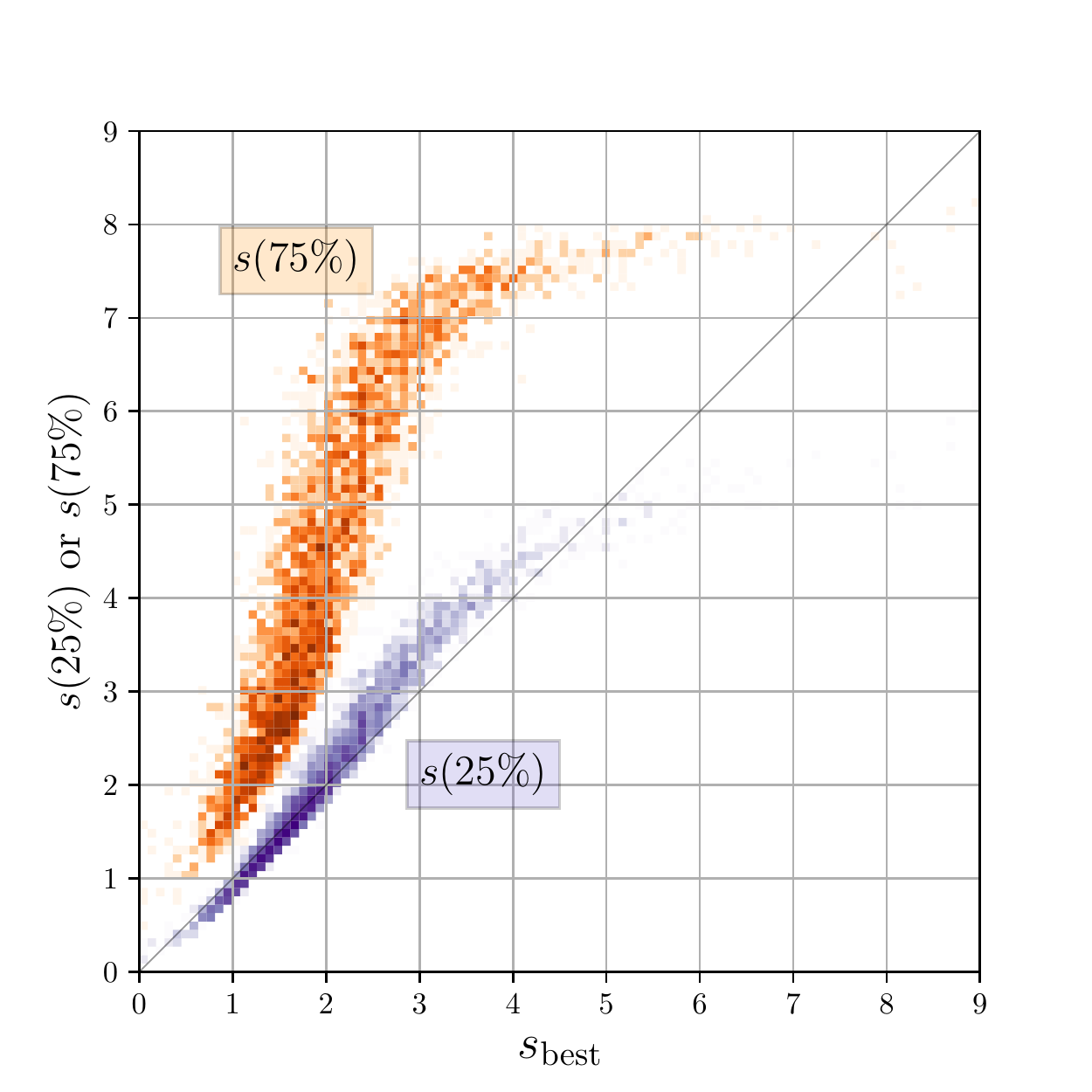}
  \caption{25\% and 75\% quantiles of MCMC $s$-distributions for well-fitted halos.
    Horizontal axis denotes the best-fit $s$-values.
    The solid line is plotted for reference, indicating the linear relation, $s_{\rm best}=s(25\%)$ or $s(75\%)$. Due to the positively skewed posterior distribution, the $s(25\%)$ is larger than $s_{\rm best}$ for most of the halos.}
  \label{fig:s_error}
\end{figure}
%%%%%%%%%%%%%%%%%%%%%%%%%%%%%%%%%%%%%%%%%%%%%%%%%%%%%%%%%%%%%%%%%%%%%%%%%%%%%%%%%
%%%%%%%%%%%%%%%%%%%%%%%%%%%%%%%%%%%%%%%%%%%%%%%%%%%%%%%%%%%%%%%%%%%%%%%%%%%%%%%%%

First look at the distribution of the parameter $s_{\rm best}$. Fig.~\ref{fig:xs} shows the distribution of the halos in our sample after the selection projected on the $s_{\rm best}$--reduced $\chi^2$ plane. We here divide the samples into two subsamples with the halo mass larger than (upper) and less than (lower) $10^{14}\,M_\odot$.
Fig.~\ref{fig:xs} shows a clear trend that the massive halos tend to have smaller reduced $\chi^2$. That is, the multi-stream structure in the massive halos is better described by the self-similar solution. A part of the reason may be that massive halos are not so severely affected by the outer environment, where merger event and asymmetric matter accretion occur. A closer look at the distribution of $s_{\rm best}$ suggests that a larger value of $s_{\rm best}$ is generally favored for massive halos. Looking at the uncertainty in the parameter estimation, however, this is not statistically significant. 

In Fig.~\ref{fig:s_error}, the $25\%$ and $75\%$ quantiles of the posterior distribution of $s$ are evaluated in each halo from the MCMC analysis, and the results are plotted as function of the best-fit value, $s_{\rm best}$. As we have seen in Fig.~\ref{fig:halo_00019_mcmc_wop}, the posterior distribution of $s$ is largely skewed with a long tail. This trend is generally seen in most of the halos in the selected sample, and the size of the $1\sigma$ error, $\Delta s$, is almost the same as $s_{\rm best}$, i.e., $\Delta s/s_{\rm best}\sim1$. Note that as increasing $s_{\rm best}$, the distribution of $75\%$ quantile apparently converges to $7-8$.  This might be partly ascribed to our setup of the prior $s\in[0,9]$, but the number of halos having $s(76\%)\sim8$ is actually small, and it does not affect the best-fit values of $s$ at least for the selected halo samples. In any case, with a large scatter in the posterior distribution, the only thing that one can clearly say from the distribution of $s_\mathrm{best}$ is that the accretion rate parameter lies at $1\lesssim s\lesssim 3$ for the selected halo samples, and there is no statistically significant difference between massive and less massive halos. 

\begin{figure}
  \centering
  \includegraphics[clip,height=56truemm]{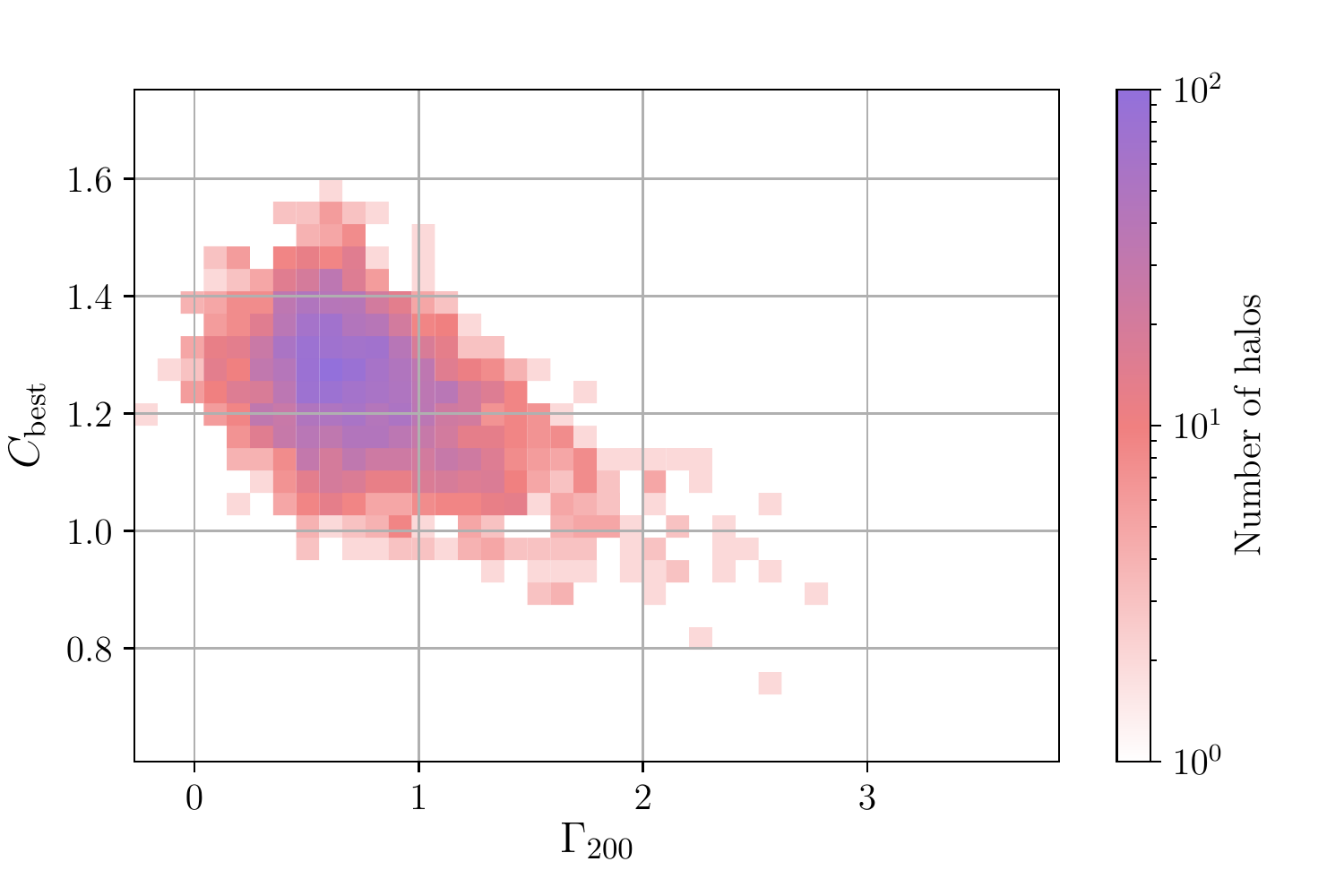}
  \includegraphics[clip,height=56truemm]{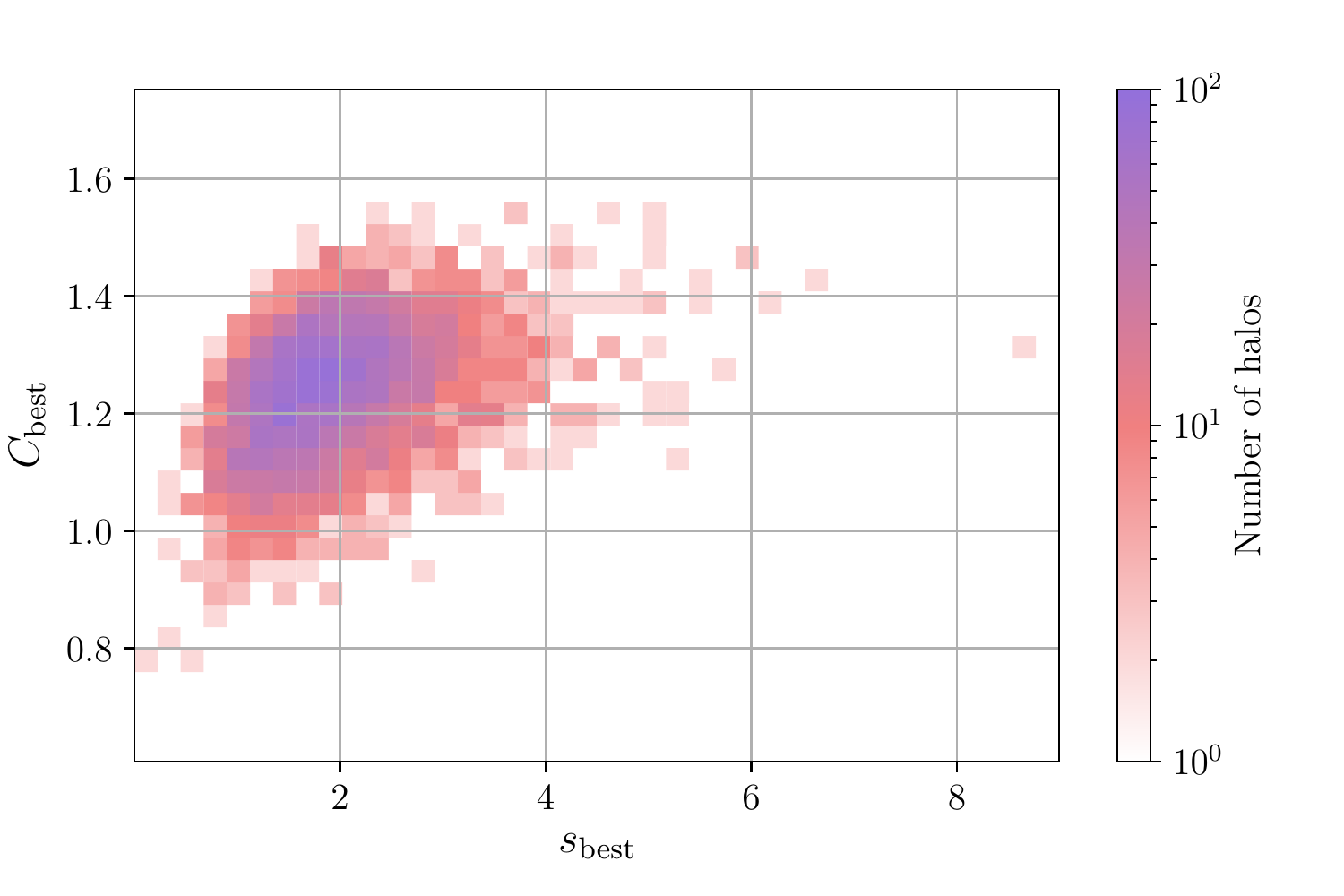}
  \caption{Correlation among the best-fit parameters in self-similar solution and measured quantity. Top panel shows the correlation between the best-fit parameter of $C$ (i.e., $C_{\rm best}$) and the measured accretion rate $\Gamma_{200}$, while bottom panels presents the result between best-fit values $C_{\rm best}$ and $s_{\rm best}$. Color depth in each pixel indicates the number of halos falling into the pixel.}
  \label{fig:c}
\end{figure}

Next look at the statistical correlation between the measured halo properties and the best-fit parameters in self-similar solution. Fig.~\ref{fig:c} show the distribution in the plane of $C_{\rm best}$ and $\Gamma_{200}$ (top) and $C_{\rm best}$ and $s_{\rm best}$ (bottom). Since $\Gamma_{200}$ and $s_\mathrm{best}$ are expected to characterize roughly the same thing, i.e., the accretion rate, a naive expectation is that these two panels exhibit a similar trend. However, the resultant correlation properties are rather different. While the measured accretion rate $\Gamma_{200}$ exhibits an anti-correlation with $C_{\rm best}$, the best-fit accretion rate $s_{\rm best}$ looks a weak but positive correlation. Recalling the fact that $C_{\rm best}$ corresponds to $R_{\rm sp}/R_{200}$, the former trend is pretty much consistent with those found in the literature \citep[e.g.,][]{mdk15,sparta2}. 

On the other hand, the weakly positive correlation between $C_{\rm best}$ and $s_{\rm best}$ looks bit puzzling, and seems to contradict with theoretical prediction by \citet{shi16}, who has derived the analytical relation between the accretion rate and splashback radius based on the self-similar solution. A large difference between \citet{shi16} and our analysis is that we treat these parameters free to be determined by fitting the measured phase-space structures to the self-similar solutions. As we discussed in Sec.~\ref{subsec: fitting method}, the parameters $C$ and $s$ are tightly related with each other in an idealistic situation, and  \citet{shi16} actually used this to derive the correlation property in an analytical way. Thus, the results shown in Fig.~\ref{fig:c} suggests a departure from the idealistic situation in the simulated halo samples. In this respect, the fitted values of the accretion rate parameter, $s_{\rm best}$, may not necessarily correspond to the net accretion rate measured at $r_{200}$, $\Gamma_{200}$.

\begin{figure}
  \centering
  \includegraphics[clip,width=80truemm]{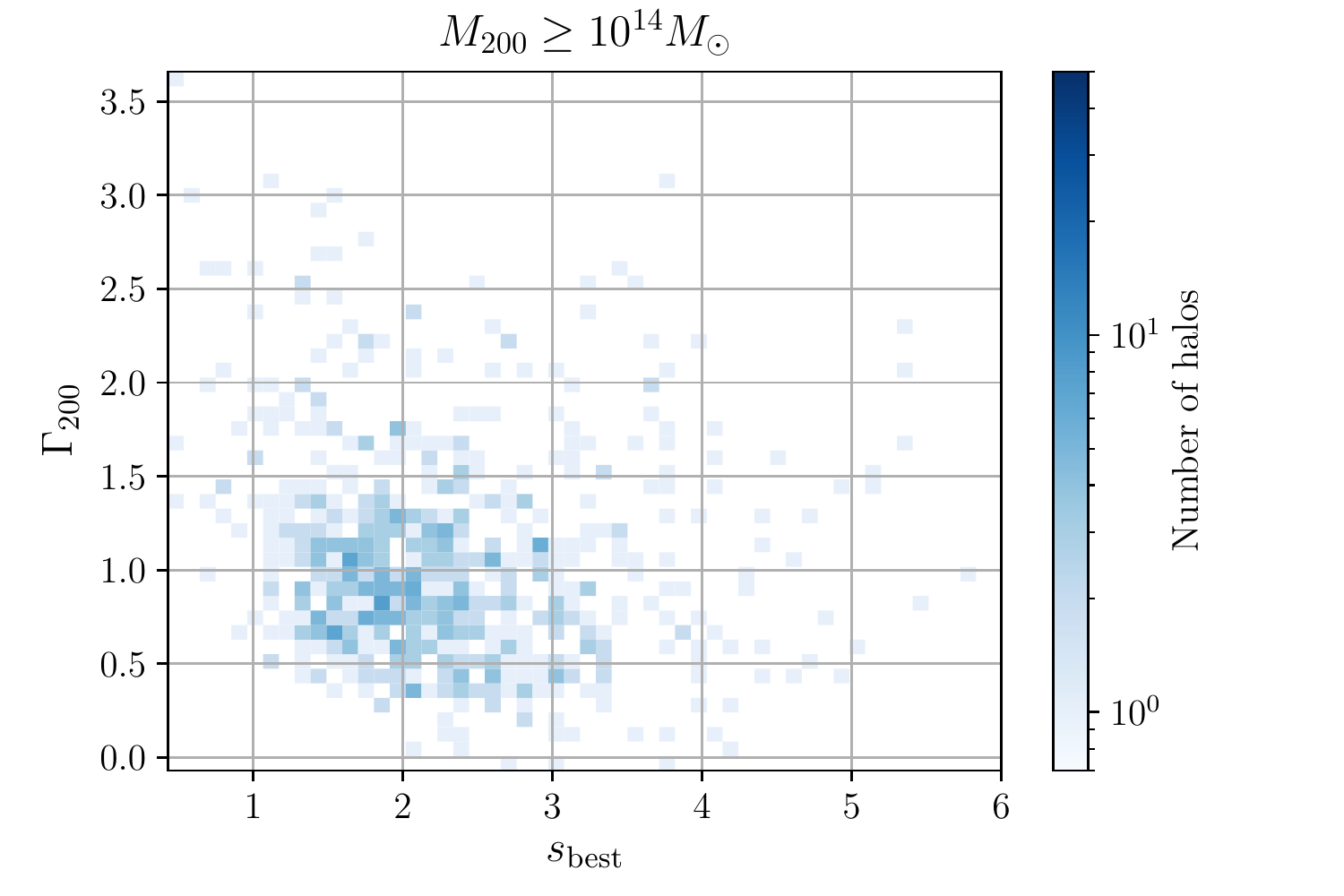}
  \includegraphics[clip,width=80truemm]{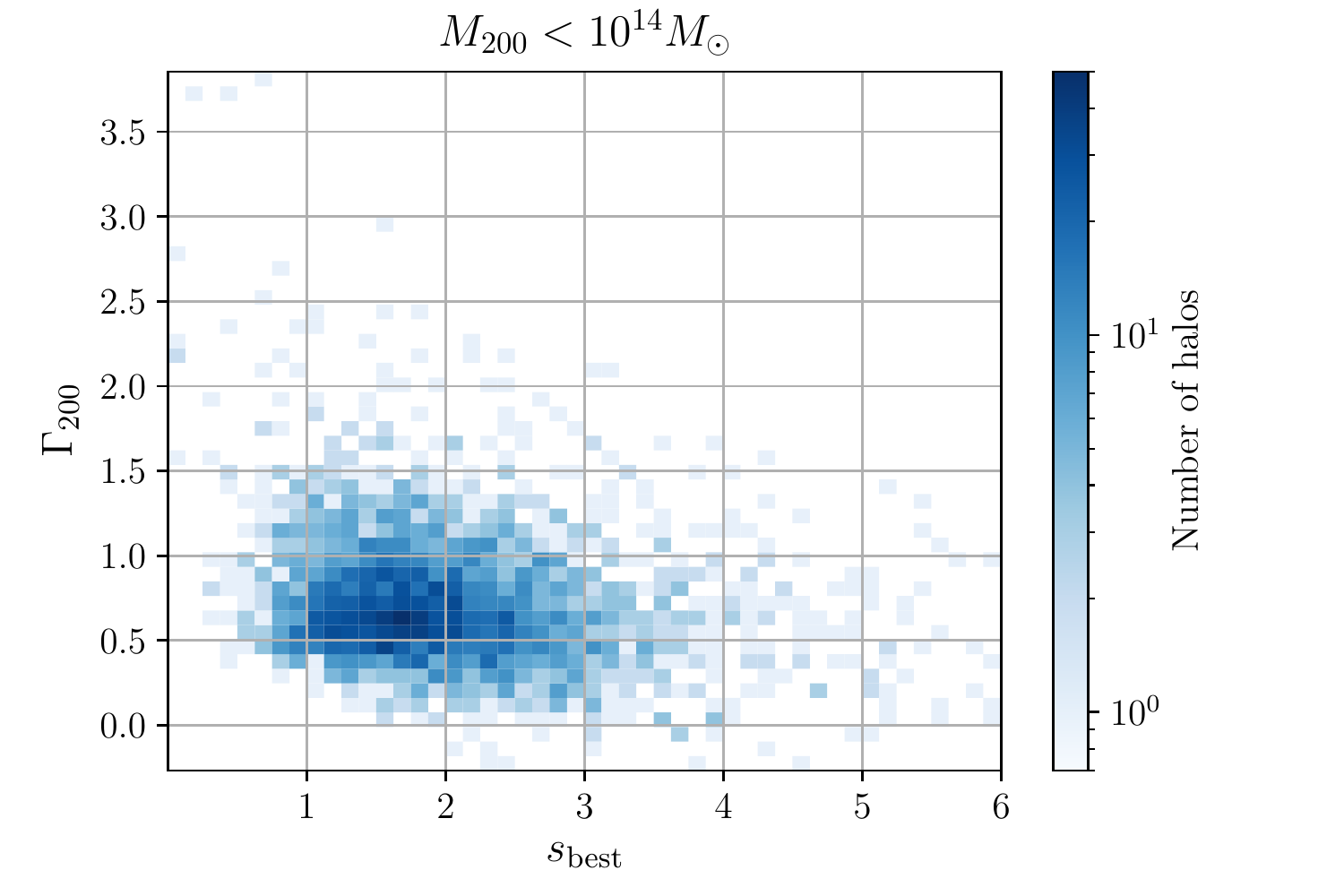}
  \caption{Correlation between the best-fit value of $s$ and measured accretion rate $\Gamma_{200}$ for well-fitted halo samples. Top and bottom panels respectively show the results for halos greater and less than mass $M_{200}=10^{14}\,M_\odot$. Color depth indicates the number of halos in each pixel.}
  \label{fig:s}
\end{figure}

To see it more explicitly, we plot in Fig.~\ref{fig:s} the statistical correlation between $s_{\rm best}$ and $\Gamma_{200}$. Here, dividing the selected halo samples into two subsamples with mass larger than (upper) and less than (lower) $10^{14}\,M_\odot$, the frequency distributions of halos are shown in the two-dimensional plane. As anticipated, there is little correlation between $s_\mathrm{best}$ and $\Gamma_{200}$, and the trend is almost similar between light and heavy subsamples, although the scatter is relatively large for massive halos. The result indicates that the two parameters are probing different aspects of the halo accretion history.

%%%%%%%%%%%%%%%%%%%%%%%%%%%%%%%%%%%%%%%%%%%%%%%%%%%%%%%%%%%%%%%%%%%%%%%%%%%%%%%%%
%%%%%%%%%%%%%%%%%%%%%%%%%%%%%%%%%%%%%%%%%%%%%%%%%%%%%%%%%%%%%%%%%%%%%%%%%%%%%%%%%
%%%%%%%%%%%%%%%%%%%%%%%%%%%%%%%%%%%%%%%%%%%%%%%%%%%%%%%%%%%%%%%%%%%%%%%%%%%%%%%%%
%%%%%%%%%%%%%%%%%%%%%%%%%%%%%%%%%%%%%%%%%%%%%%%%%%%%%%%%%%%%%%%%%%%%%%%%%%%%%%%%%

%%%%%%%%%%%%%%%%%%%%%%%%%%%%%%%%%%%%%%%%%%%%%%%%%%%%%%%%%%%%%%%%%%%%%%%%%%%%%%%%%
%%%%%%%%%%%%%%%%%%%%%%%%%%%%%%%%%%%%%%%%%%%%%%%%%%%%%%%%%%%%%%%%%%%%%%%%%%%%%%%%%
%%%%%%%%%%%%%%%%%%%%%%%%%%%%%%%%%%%%%%%%%%%%%%%%%%%%%%%%%%%%%%%%%%%%%%%%%%%%%%%%%
%%%%%%%%%%%%%%%%%%%%%%%%%%%%%%%%%%%%%%%%%%%%%%%%%%%%%%%%%%%%%%%%%%%%%%%%%%%%%%%%%

%%%%%%%%%%%%%%%%%%%%%%%%%%%%%%%%%%%%%%%%%%%%%%%%%%%%%%%%%
%%%%%%%%%%%%%%%%%%%%%%%%%%%%%%%%%%%%%%%%%%%%%%%%%%%%%%%%%
\section{Discussion} 
\label{sec:discussion}
%%%%%%%%%%%%%%%%%%%%%%%%%%%%%%%%%%%%%%%%%%%%%%%%%%%%%%%%%
%%%%%%%%%%%%%%%%%%%%%%%%%%%%%%%%%%%%%%%%%%%%%%%%%%%%%%%%%

In this section, to better understand the results shown in Figs.~\ref{fig:c} and \ref{fig:s}, 
we investigate the physical meaning of the parameter $s_\mathrm{best}$, and look for a link to other quantities measured from $N$-body simulations. For this purpose, we decompose the mass of each halo into different contributions, each of which consists of DM particles with different numbers of apocenter passages. Then, we consider the contribution coming from the DM particles having $p \geq p_\mathrm{min}$ at redshift $z$. Denoting the mass of such a contribution by $M_{p \geq p_\mathrm{min}}( z )$, we define the new accretion rate parameters, which should be more relevant to the multi-stream flows inside the splashback radius as:
%%%%%%%%%%%%%%%%%%%%%%%%%%%%%%%%%%%%%%%%%%%%%%%%%%%%%%%%%%%%%%%%%%%%%%%%%%%%%%%%%
\begin{align}
  \Gamma_{p \geq p_\mathrm{min}} = \frac{ \Delta \ln M_{p \geq p_\mathrm{min}} }{ \Delta \ln a }.
  \label{eq:Gppmin}
\end{align}
%%%%%%%%%%%%%%%%%%%%%%%%%%%%%%%%%%%%%%%%%%%%%%%%%%%%%%%%%%%%%%%%%%%%%%%%%%%%%%%%%
In evaluating Eq.~(\ref{eq:Gppmin}),  the finite difference, $\Delta \ln M_{p \geq p_\mathrm{min}}$, is taken between $z = 0$ and $z = 0.11$, not $z=0$ and $z=0.5$, which we adopted in measuring $\Gamma_{200}$ [see Eq.~(\ref{eq:Gamma200})]. The reason is that increasing $z$ as well as $p$, a reliable estimation of the number of apocenter passages becomes difficult due to the limited range of available redshifts ($z\leq 1.43$ in our case). The closer redshift interval used in Eq.~(\ref{eq:Gppmin}) gives us a more instantaneous estimate of the mass accretion.

\begin{figure}
  \centering
  \includegraphics[clip,width=63truemm]{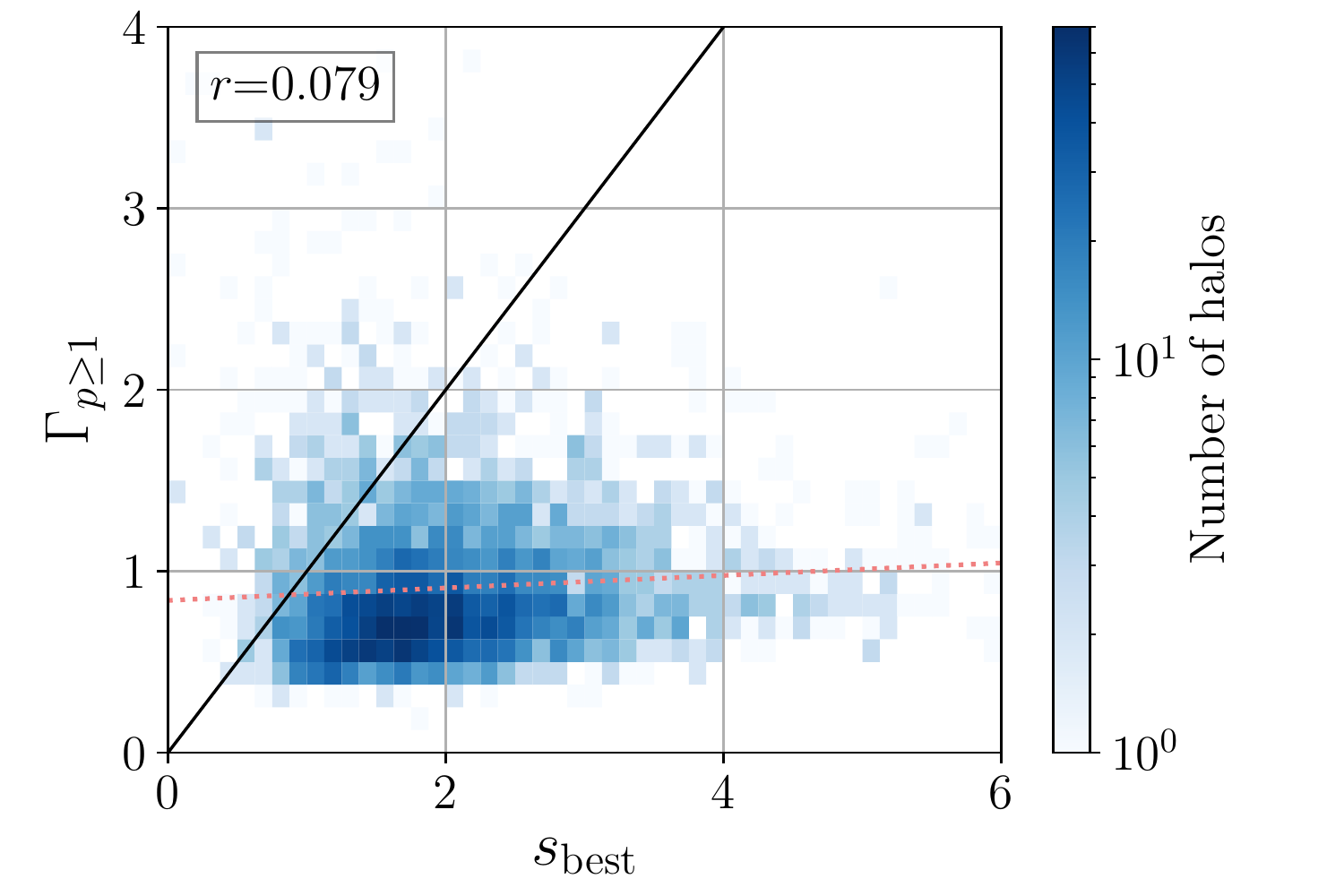}
  \includegraphics[clip,width=63truemm]{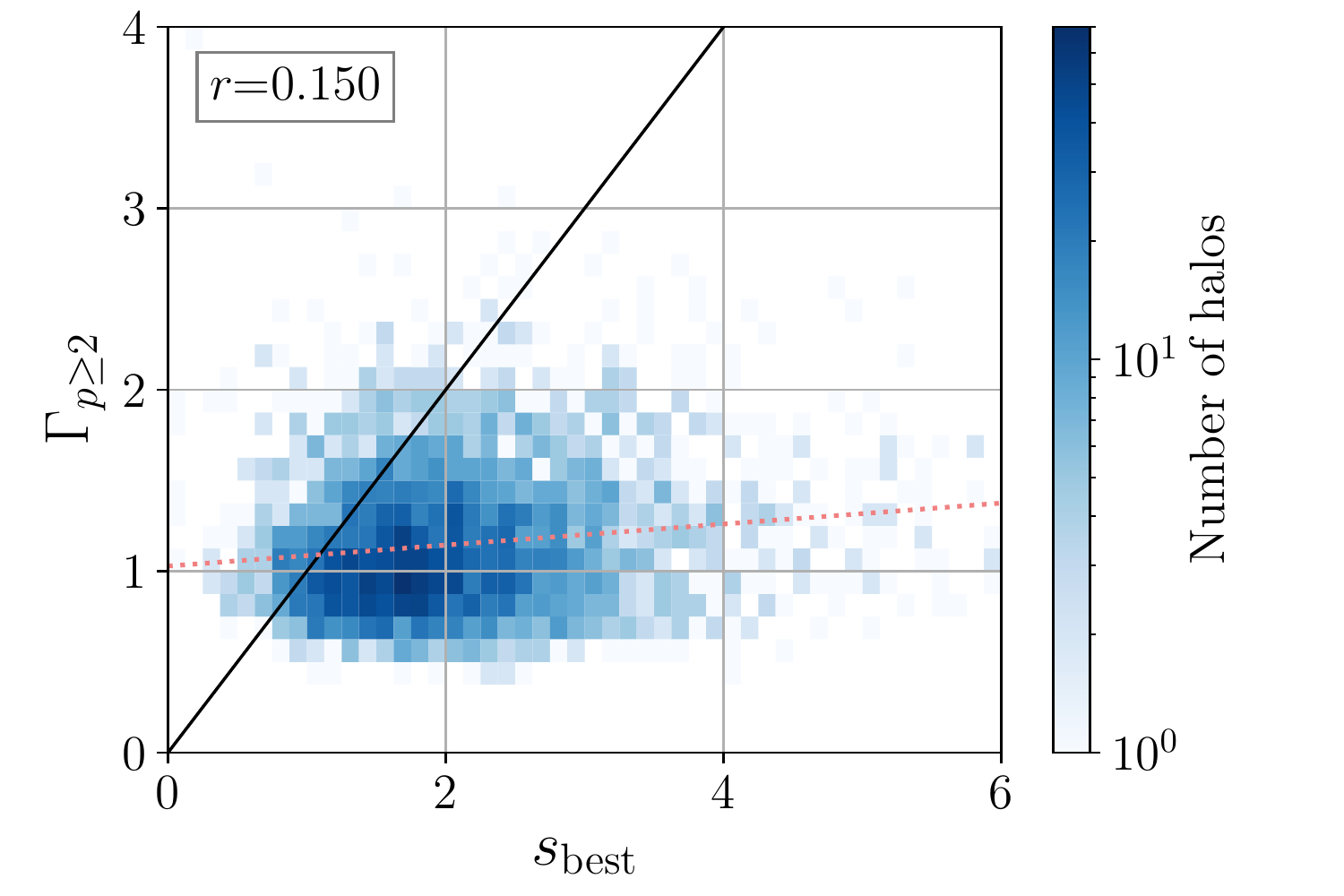}
  \includegraphics[clip,width=63truemm]{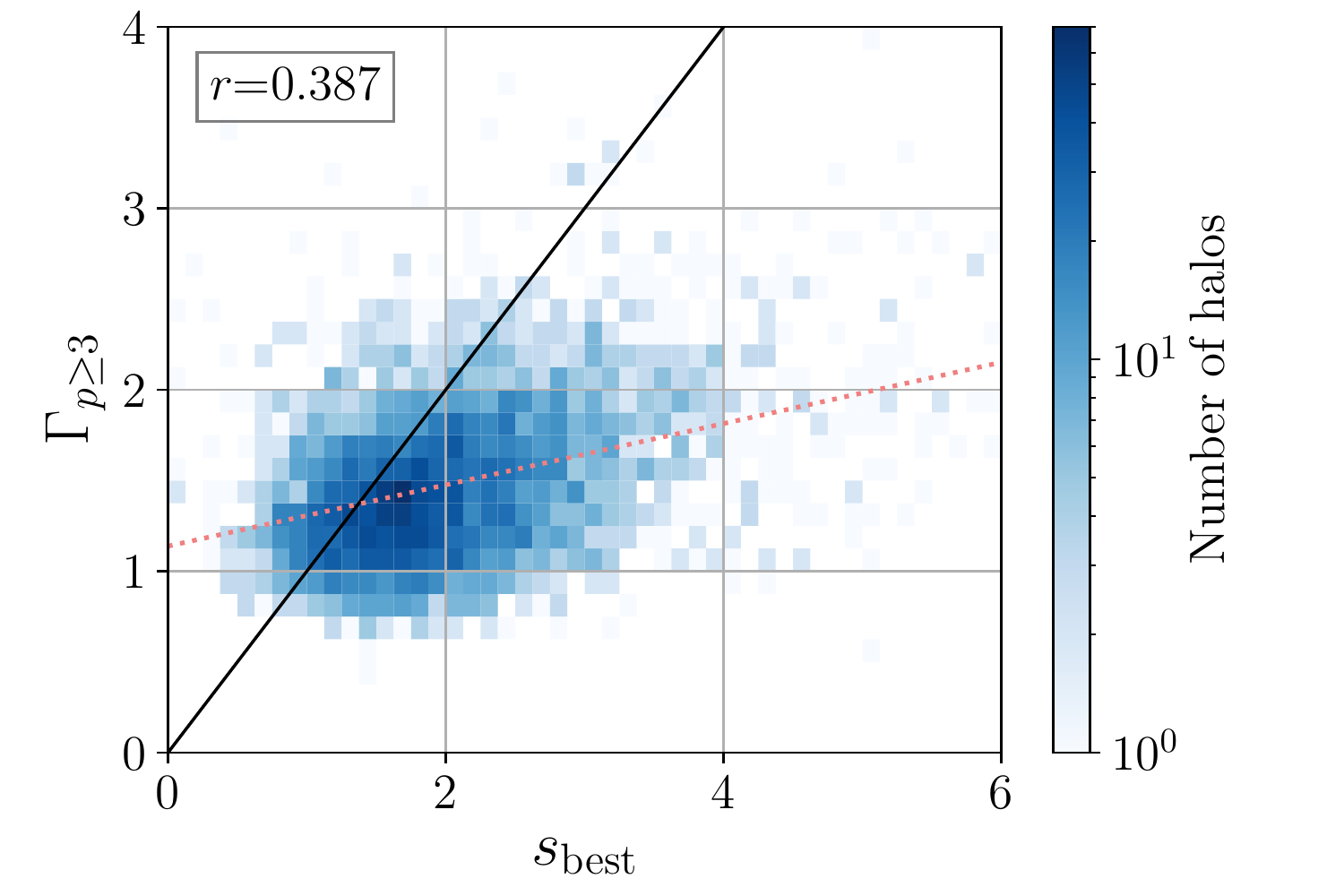}
  \includegraphics[clip,width=63truemm]{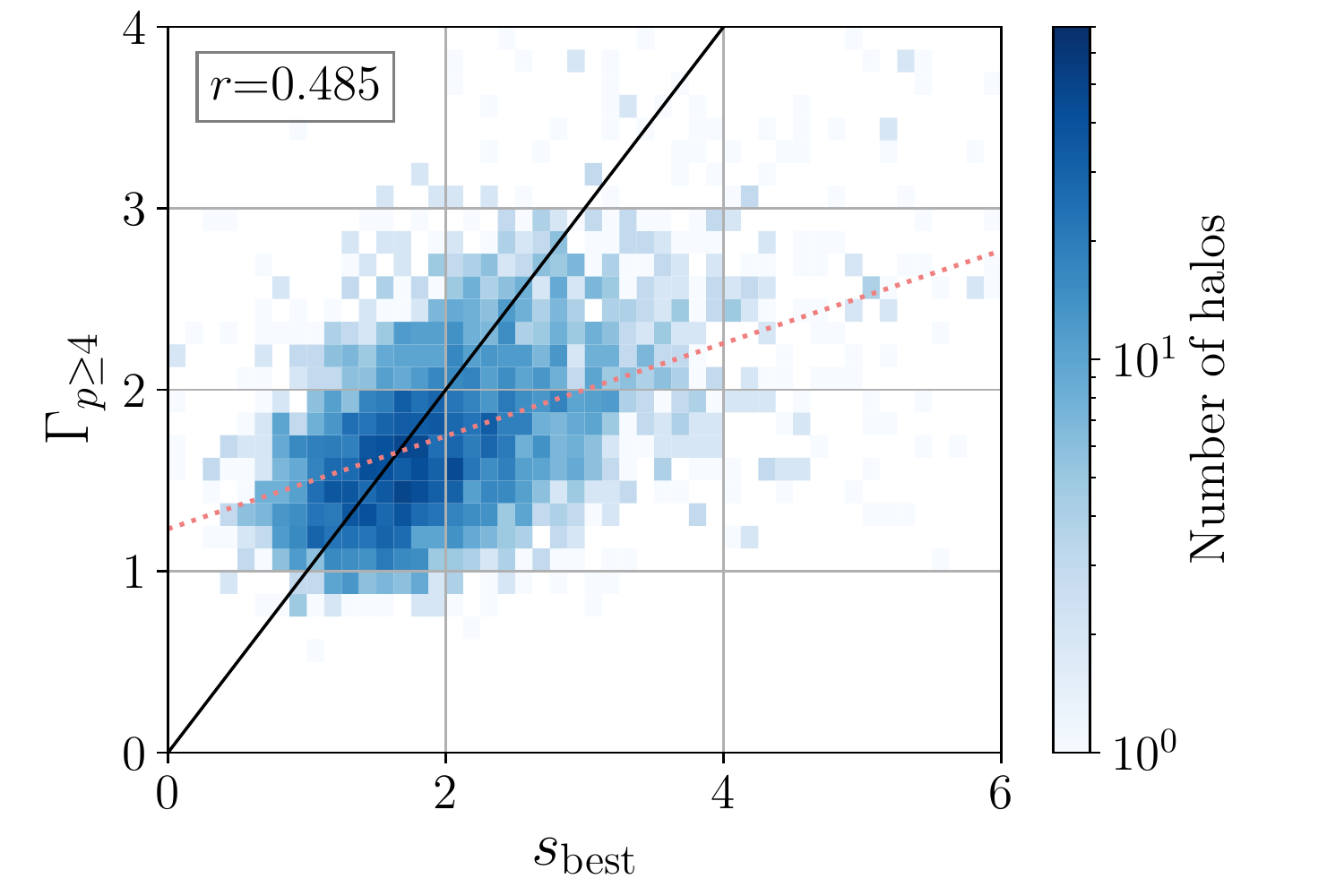}
  \includegraphics[clip,width=63truemm]{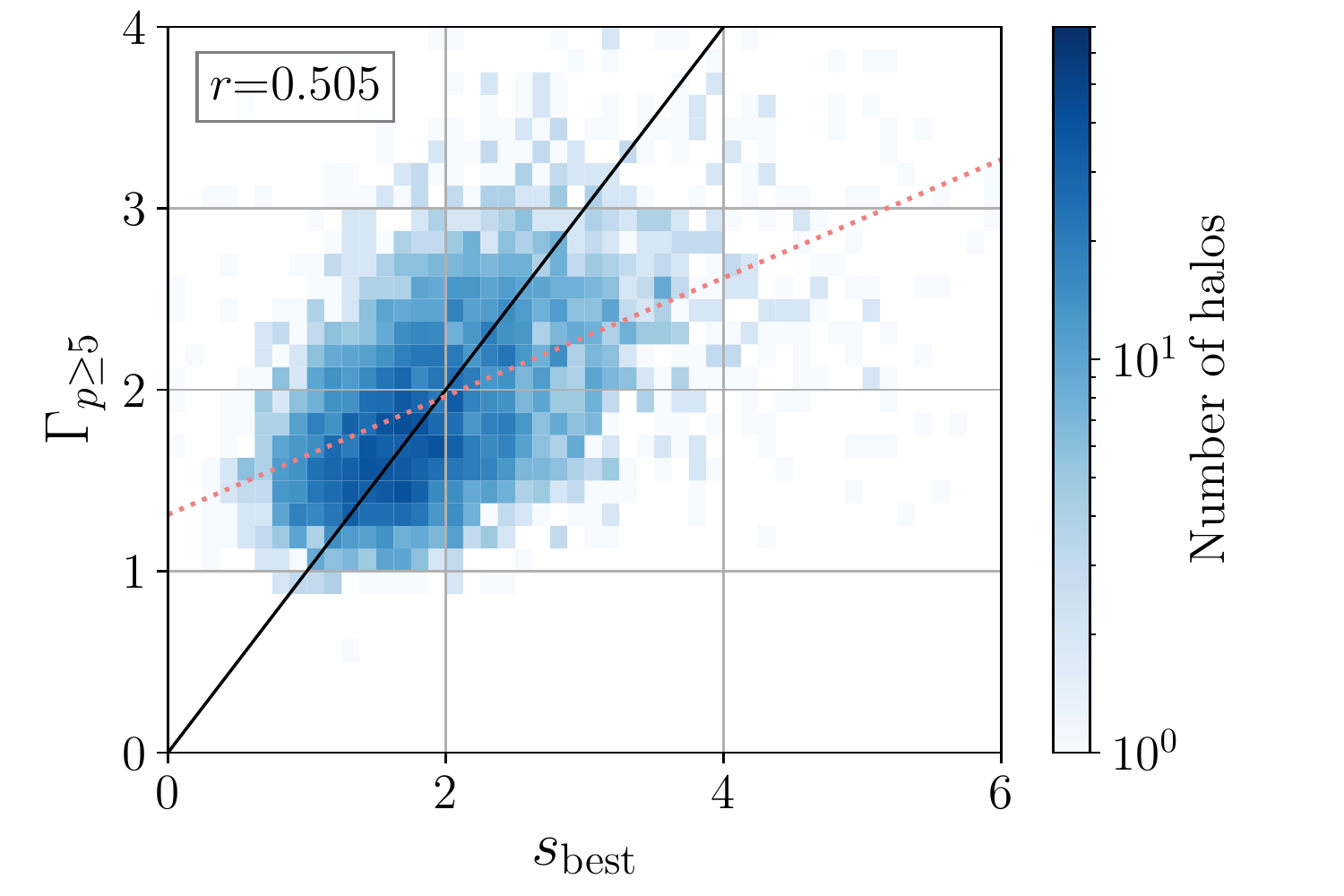}
  \caption{Correlation between $s_\mathrm{best}$ and $\Gamma_{p \geq p_\mathrm{min}}$ defined at Eq.~(\ref{eq:Gppmin}) for well-fitted halo samples. From top to bottom, the results with $p_\mathrm{min}=1$ to $5$ are respectively shown. Color depth in each pixel indicates the number of halos in logarithmic scales. Dotted lines are the linear regression estimated by least squares method, and the derived values of Pearson's correlation coefficient, denoted by $r$, are also shown in each panel. For reference, linear relations of $s_\mathrm{best} = \Gamma_{p \geq p_\mathrm{min}}$ are also plotted in black solid lines. }
  \label{fig:pq}
\end{figure}

Fig.~\ref{fig:pq} shows the correlations between $s_\mathrm{best}$ and $\Gamma_{p \geq p_\mathrm{min}}$ for $p_{\rm min}=1-5$ (from top to bottom). In each case, we perform linear regression and plot the result by the red dotted line. Also, we estimate the Pearson correlation coefficient $r$, and the derived values are shown in each panel. For reference, the linear relation of $s_{\rm best}=\Gamma_{p\geq p_{\rm min}}$ is also plotted in black solid line.  As increasing $p_{\rm min}$, the estimated values of $\Gamma_{p \geq p_\mathrm{min}}$ gets large and exceeds the mean value of $\Gamma_{200}$ (roughly $\sim1$). This is presumably because we are preferentially looking at the inner halo structures, where the inward streaming flows become dominant. By contrast, the inward accretion flow near the halo boundary is prone to be disturbed by the outer environment, and hence $\Gamma_{p \geq p_\mathrm{min}}$ tends to get larger than $\Gamma_{200}$. A notably interesting trend we find is that the correlation between $s_{\rm best}$ and $\Gamma_{p\geq p_{\rm min}}$ gets tighter as increasing the minimum number of apocenter-passages, $p_\mathrm{min}$. The trend is, indeed, more clearly seen in the Pearson correlation coefficient, and quantitatively, the coefficient increases from $0.079$ to $0.505$ as we change $p_{\rm min}$ from $1$ to $5$. Further,  the slope of the linear regression gets inclined, approaching to $s_{\rm best}=\Gamma_{p \geq p_\mathrm{min}}$, though the regression coefficient is still $0.67$ at $p_{\rm min} = 5$. The result suggests that the parameter $s_\mathrm{best}$ is determined by the inner multi-stream flows with a large value of $p$. In other words, the best-fit value of $s$ carries some information on the memories of the early-phase mass accretion history.  By contrast, the accretion rate $\Gamma_{200}$ is sensitive to the recently accreting matter near the halo boundary.  In this sense, little correlation between $s_\mathrm{best}$ and $\Gamma_{200}$, shown in Fig.~\ref{fig:s}, may be regarded as a reasonable outcome.

Finally, it would be interesting if the quantity similar to $s_{\rm best}$ can be measured directly from observations. The complementarity of the parameter $s$ to $\Gamma_{200}$ gives a fruitful insight into the history of the halo formation and evolution over a longer period of time. Due to the fact that we can measure only the line-of-sight component of the velocity and/or confusion between Hubble flow and peculiar velocity, it is not actually straightforward to get access to the phase-space structure \citep[but see e.g.,][]{clash,mbm14,galweight}. Nevertheless, as we have seen in Sec.~\ref{sec:self-similar}, the parameter $s$ is related to the inner slope of a halo [Eq.~(\ref{eq:asymptotic_slope})]. Although it is indirect, the density slope could provide a useful hint to infer or pin down the early-phase mass accretion history of a halo. Other proxies, such as the color or morphological information for galaxies, might be useful to infer the streams in the phase space.

%%%%%%%%%%%%%%%%%%%%%%%%%%%%%%%%%%%%%%%%%%%%%%%%%%%%%%%%%%%%%%%%%%%%%%%%%%%%%%%%%
%%%%%%%%%%%%%%%%%%%%%%%%%%%%%%%%%%%%%%%%%%%%%%%%%%%%%%%%%%%%%%%%%%%%%%%%%%%%%%%%%
%%%%%%%%%%%%%%%%%%%%%%%%%%%%%%%%%%%%%%%%%%%%%%%%%%%%%%%%%%%%%%%%%%%%%%%%%%%%%%%%%
%%%%%%%%%%%%%%%%%%%%%%%%%%%%%%%%%%%%%%%%%%%%%%%%%%%%%%%%%%%%%%%%%%%%%%%%%%%%%%%%%

%%%%%%%%%%%%%%%%%%%%%%%%%%%%%%%%%%%%%%%%%%%%%%%%%%%%%%%%%%%%%%%%%%%%%%%%%%%%%%%%%
%%%%%%%%%%%%%%%%%%%%%%%%%%%%%%%%%%%%%%%%%%%%%%%%%%%%%%%%%%%%%%%%%%%%%%%%%%%%%%%%%
\section{Conclusions}
\label{sec:conclusion}
%%%%%%%%%%%%%%%%%%%%%%%%%%%%%%%%%%%%%%%%%%%%%%%%%%%%%%%%%%%%%%%%%%%%%%%%%%%%%%%%%
%%%%%%%%%%%%%%%%%%%%%%%%%%%%%%%%%%%%%%%%%%%%%%%%%%%%%%%%%%%%%%%%%%%%%%%%%%%%%%%%%

In this paper, we have studied the radial phase-space properties of cold dark matter halos in a cosmological $N$-body simulation. In particular, we have quantified the multi-stream structures of halos inside the splashback radius, and their radial phase-space distributions are compared with the spherically symmetric self-similar solution by \citet{fg84}. In order to trace and characterize the multi-stream nature of each halo in $N$-body simulation, we implemented the SPARTA algorithm developed by \citet{sparta1} to keep track of the trajectories of dark matter particles. We extended it to identify the inner apocenter passages inside the so-called splashback radius, and count its number along each trajectory of dark matter particle. With the particle distribution characterized by the number of apocenter passages, the multi-stream nature of dark matter velocity flows can be visualized in phase space, and we were able to make a detailed comparison of the phase-space properties with the predictions of the self-similar solution. Using Markov-chain Monte Carlo technique, we have analyzed in total $11,296$ halos with mass $M_{200}\geq10^{13}\,M_\odot$ to obtain the best-fit parameters of the self-similar solution characterizing the multi-stream flows inside the halos in $N$-body simulation.

Our important findings are summarized as follows:
\begin{itemize}
    \item About $30\%$ of the halos among those we analyzed are classified as well described by the self-similar solution \citep{fg84}. These halos are selected by imposing the three conditions discussed in Sec.~\ref{subsec:sample_selection}, i.e., (i) sufficient number of particles in most of the radial-velocity bins, (ii) a clear determination of stream line/shell tagged with the number of apocenter passages, (iii) a condition for the goodness-of-fit for each stream given by Eq.~(\ref{chi2_p}) (see also Table \ref{tb:halo_selection}). Typical examples of well-fitted halos are shown in Figs.~\ref{fig:halo_00019} and \ref{fig:halo_01244}. We found that more massive halos tend to be better described by the self-similar solution with a smaller value of the reduced $\chi^2$ (see Fig.~\ref{fig:xs}).
\\
    \item The self-similar solution by \citet{fg84} is characterized by the three parameters: stationary accretion rate $s$, and scaling parameters in radial position and velocity, $C$ and $U$, where the parameters $C$ is related to the ratio of splashback to virial radius through $C=R_{\rm sp}/R_{200}$. Allowing these parameters to be free, we determined their best-fit values in each halo, and found that for the well-fitted halo sample, the best-fit values of $s$ and $C$ are distributed around the ranges $1\lesssim s_{\rm best}\lesssim3$ and $0.9\lesssim C_{\rm best} \lesssim1.5$ (see Fig.~\ref{fig:c}).
 \\
    \item Statistical analysis of the well-fitted halo sample reveals that the best-fit model parameter $C_{\rm best}$ show an anti-correlation with the measured accretion rate at $R_{200}$, $\Gamma_{200}$ [see Eq.~(\ref{eq:Gamma200})]. While this is fully consistent with those previously found in the literature, the parameter $C_{\rm best}$ exhibits a weak but positive correlation with the best-fit accretion rate parameter, $s_{\rm best}$, which apparently contradicts with previous findings. In particular, we found that there is no clear correlation between $s_{\rm best}$ and $\Gamma_{200}$.  A detailed study on the mass accretion rate (Sec.~\ref{sec:discussion}) indicates that the best-fit parameter $s_{\rm best}$ in the self-similar solution rather characterizes the accretion rate determined by the inner structure of halos with a large value of $p$ (number of apocenter passage). In other words, $s_{\rm best}$ is the quantity complementary to $\Gamma_{200}$ and carries the information on the early-phase mass accretion history, also linked to the slope of density profile inside the splashback radius. 
\end{itemize}

Note that these findings are based on an $N$-body simulation performed in an Einstein-de Sitter cosmology. One obvious question is whether these behaviors persist in standard $\Lambda$CDM cosmology or not. Although we lose strict self-similarity, recalling the fact that dynamical time-scale of halo formation is shorter than the time-scale of cosmic expansion, one expects that the similar features can be still seen, especially at the inner streams in massive halos formed at an early time. In fact,  with a slight extension of the self-similar solution, the analytical relation derived by \citet{shi16} is found to describe the $N$-body halos well in a $\Lambda$CDM cosmology \citep{sparta2}. In any case, a quantitative study on the radial phase-space structure of halos in non-Einstein-de Sitter Universe is worth for further investigation, and we will address this issue in near future. 

From the observational point-of-view, a more crucial and interesting aspect to be clarified would be the phase-space structure of subhalos and satellite galaxies inside a halo in connection with dark matter multi-stream flows. These objects are known to be affected by dynamical friction, and because of this, their splashback features are systematically different from that of the dark matter \citep{Adhikari_etal2016,adhikari18}. In this respect, their phase-space distribution would not exactly trace the multi-stream structure of dark matter. Characterising and modeling their phase-space properties are important for confronting observations. For this purpose, a systematic study using high-resolution cosmological simulations with a large boxsize is indispensable, and it may even give a hint to probe the nature of cold dark matter from observations. 

Finally, the phase-space study of DM halos definitely provides an important and new clue to understand the physical properties of CDM halos. Recently, alternative to the $N$-body simulation, a more fundamental numerical method directly solving Vlasov-Poisson equations in $6D$ phase space is developing \citep{Yoshikawa_etal2013,Hahn_Angulo2016,Sousbie_Colombi2016}. Such a technique would be certainly essential to resolve the phase-space structures deep inside the splashback, thus giving a hint to clarify the nature of cuspy density profile as well as the power-law nature of pseudo phase-space density profile. Although the present paper focuses on the phase-space structure inside the splashback radius, it is still regarded as the outer part of the halo system away from the cuspy structure. Nevertheless, an extension of the analysis to the inner phase-space is straightforward. The investigation of the self-similarity and the structural properties is a crucial step toward the understanding of generic features of CDM halos, and this is left for future work.

\section*{Acknowledgments}
This work was initiated during the invitation program of JSPS Grant No. L16519. This work was granted access to HPC resources of CINES through allocations made by GENCI (Grand Equipement National de Calcul Intensif) under the allocations 2017-A0010402287 and 2018-A0030402287. Numerical computation was also carried out partly at the Yukawa Institute Computer Facility. YR thanks Shankar Agarwal for fruitful discussions about halo definitions. AT and HS would like to thank St\'ephane Colombi for useful comments  and references. We thank Benedikt Diemer for useful comments. This work was supported in part by MEXT/JSPS KAKENHI Grant Numbers JP15H05889, JP16H03977 (AT), JP17K14273, and JP19H00677 (TN), and Japan Science and Technology Agency CREST JPMHCR1414 (TN). 

%%%%%%%%%%%%%%%%%%%%%%%%%%%%%%%%%%%%%%%%%%%%%%%%%%

%%%%%%%%%%%%%%%%%%%% REFERENCES %%%%%%%%%%%%%%%%%%
\bibliographystyle{mnras}
\bibliography{reference}
%%%%%%%%%%%%%%%%%%%%%%%%%%%%%%%%%%%%%%%%%%%%%%%%%%

%%%%%%%%%%%%%%%%% APPENDICES %%%%%%%%%%%%%%%%%%%%%
%%%%%%%%%%%%%%%%%%%%%%%%%%%%%%%%%%%%%%%%%%%%%%%%%%

\end{document}